\documentclass[a4paper,11pt]{article}
\pdfoutput=1 % if your are submitting a pdflatex (i.e. if you have
\usepackage{aas_macros}
\usepackage{jcappub} 
\usepackage[T1]{fontenc} % if needed
\usepackage{amsbsy}
\usepackage{color}
\usepackage{xcolor}
\usepackage{subcaption}
\usepackage{hyperref}

\usepackage{multicol}

\usepackage[normalem]{ulem}

\newcommand{\reversedpm}{\mathrel{\rlap{\raisebox{0.95ex}{\scalebox{0.7}{$-$}}}\raisebox{0.15ex}{\scalebox{0.7}{$+$}}}}

\newcommand{\be}{\begin{equation}}
\newcommand{\ee}{\end{equation}}

\usepackage{amssymb}

\title{\boldmath From inflation to dark matter halo profiles: the impact of primordial non-Gaussianities on the central density cusp}

\author[a]{Cl\'ement Stahl}
\author[a]{Nicolas Mai}
\author[a]{Benoit Famaey}
\author[b]{Yohan Dubois}
\author[a]{Rodrigo Ibata}

\affiliation[a]{Universit\'e de Strasbourg, CNRS, Observatoire astronomique de Strasbourg, UMR 7550, 67000 Strasbourg, France}
\affiliation[b]{Institut d'Astrophysique de Paris, CNRS, Sorbonne Universit\'{e}, UMR 7095, 98bis bd Arago, 75014 Paris, France}

\emailAdd{clement.stahl@unistra.fr}

\abstract{It has recently been shown that local primordial non-Gaussianities (PNG) with significant amplitude ($|f_{\rm NL}| \sim 1000$), at small (Mpc) scales, can help in forming simulated galaxies with more disky baryonic kinematics than in the Gaussian case, while generating matter power spectra that can differ by up to 20\% from the Gaussian case at non-linear scales. Here, we explore in detail the consequences of such small-scale PNG on the dark matter halo profiles. We show in particular that, for negative $f_{\rm NL}$, dark matter halos formed in collisionless simulations are not always well described by the traditional Navarro-Frenk-White (NFW) profiles, as supported by their sparsity distribution. We conclude that NFW profiles are not as clear attractors for the density profiles of dark matter halos in the presence of PNG than in the case of a Gaussian contrast density field. We show how alternatives to the NFW profile can describe halos both in the Gaussian and non-Gaussian cases. From the combination of our sparsity analysis and the quality of the adjustments of the density profiles with a minimal extension to NFW, we conclude that $z=1$ halos carry the most interesting information about PNG.}

\begin{document}
\maketitle
\flushbottom

\section{Introduction}\label{sec:Introduction}
The $\Lambda$CDM ($\Lambda$ Cold Dark Matter) standard model of cosmology relies on different assumptions: (i) gravity is described by General Relativity; (ii) the universe is composed of particles from the standard model of particles physics (including radiation and neutrinos) as well as cold dark matter (CDM) and dark energy (described by a cosmological constant $\Lambda$); (iii) the universe is statistically homogeneous and isotropic on large scales (the cosmological principle);
(iv) the universe started out, after the inflationary phase, in a nearly scale invariant, highly Gaussian state for the density fluctuations. These working hypotheses are supported by a plethora of observational evidence including the Cosmic Microwave Background (CMB), the multiple probes of large scale structure including its baryon acoustic oscillations, the abundances of light elements. Observational campaigns are awaited for the next decade to continue the quest for precision cosmology and to challenge these hypotheses. 

Despite providing an astonishingly good fit to the observational data, $\Lambda$CDM lacks fundamental grounding for its dark sector (dark matter and dark energy). Furthermore dark clouds may be gathering in the form of, e.g., tensions in the parameter inference from early and late cosmological probes \cite{Abdalla}, or the existence of extreme objects which seem to be more often observed than what $\Lambda$CDM predicts: very massive clusters \cite{ElGordo}, very early-formed galaxies \cite{Finkelstein:2013lfa,Labbe,MBK2}, very large voids \cite{DES:2021cge}. At late times, when the evolution of matter perturbations becomes highly non-linear and has to be mostly modeled with numerical methods, there is also a handful of small scale problems \citep[e.g.,][]{Bullock,2012LRR....15...10F}.
These flies in the ointment may be hints that the ingredients of $\Lambda$CDM need to be revisited. Much of the activities in this direction have been to reconsider point (i) or (ii), in particular with the attractive hope to have a more complete theory of gravity and/or a a deeper understanding of the dark sector. We have been arguing \cite{Stahl:2023ccv,Stahl:2022did} that at least some of the above mentioned problems may perhaps find a natural solution by amending hypothesis (iv). For instance, the presence of extremely massive objects \cite{Ezquiaga:2022qpw}, the $S_8$ tension \cite{Stahl:2023ccv}, the presence of early-formed objects \cite{Biagetti:2022ode}, some $z=0$ small-scale problems \cite{Stahl:2023ccv,Stahl:2022did} may be explained, though much work still needs to be carried out to obtain a full answer to all the questions that a reconsideration of initial density fluctuations raises, and to address whether they can fully, or partially, address the aforementioned problems of vanilla $\Lambda$CDM.

Inflation \cite{Mukhanov:2005sc} is a period of accelerated expansion of the universe leading to particles production due to vacuum quantum fluctuations stretched to large scales. At leading order in cosmological perturbation theory, inflationary scenarios engender a near scale invariant Gaussian distribution. The predictions of inflation are in line with current measurements of the CMB \cite{Planck:2018vyg,Tristram:2023haj}. One has not yet directly probed the primordial matter distribution at smaller\footnote{
In this paper, we define small scale as scales smaller than CMB scales: $\mathcal{O}(10)$ Mpc.} scales. Furthermore the CMB is a cosmological probe which allows one to probe the matter probability density function rather close to its mean, and is pretty uninformative about the tail of the distribution function. As such, the CMB and large-scale structure observations do not allow yet to set constraints on PNG at Mpc scales.

Beyond leading order in perturbation theory, one however expects to have non-Gaussian tails, of varying intensity depending on the detail of the microphysics of inflation. As these tails cannot be cast as an expansion around a Gaussian probability density function (PDF), a proper modeling requires in principle non-perturbative techniques, see eg.~Refs.~\cite{Chen:2018uul,Celoria:2021vjw,Gow:2022jfb,Cohen:2022clv,Hooshangi:2023kss}. As such, these models can make more probable the extreme objects of our universe \cite{Ezquiaga:2022qpw}. Further observational prospects for these non-perturbative PNG include the search for primordial black holes \cite{Carr:2023tpt}, that could also represent a fraction of dark matter \cite{Clesse:2017bsw,Green:2020jor}, the gravitational waves cosmological background \cite{Caprini:2018mtu}, or ultracompact minihalos \cite{FrancoAbellan:2023sby}. 

As that type of model also predicts large PNG on small scales, they could also leave signatures on CMB spectral $\mu$-distortions \cite{Bianchini:2022dqh}, and on every probe going from the scale of galaxy clusters down to galaxy scales, as also probed by local (perturbative) models on small scales. These include, e.g., cluster number counts
\cite{LoVerde:2007ri}, a scale-dependent deviation from the Gaussian case in the bispectrum and trispectrum \cite{Pena:2022sdg}, weak lensing \cite{Anbajagane:2023wif} or strong lensing \cite{DAloisio:2011uez} signatures, the mass function or luminosity function of high redshift galaxies \cite{Habouzit:2014hna,Sabti:2020ser,Biagetti:2022ode}, reionisation \cite{Crociani:2008dt,Chevallard:2014sxa}, or details of the internal kinematics of galaxies \cite{Avila-Reese:2003cjm,Stahl:2023ccv}. In the present paper, we will study the density profile of halos formed in simulations with and without large (small scales) local PNG. 

The inner structure (density profile) of dark matter halos is an important aspect of the analysis of cosmological data, as it connects the observed galaxy distribution to theoretical predictions of the underlying dark matter distribution \cite{Cooray:2002dia}. It is also of primordial importance for direct and indirect search of dark matter \cite{Klasen:2015uma}. Our current best fitting functions are empirical laws calibrated on observations or on numerical experiments. In collisionless simulations \cite{Angulo:2021kes}, halos that form in a cosmological context are typically well described by a universal NFW profile \cite{Navarro:1995iw,Navarro:1996gj}, regardless of the epoch or cosmological parameters. In observations, and in hydrodynamical simulations including baryons, the situation is much less clear, but the profiles are rarely NFW on galaxy scales \cite[e.g.,][]{Jo,Korsaga_2023}, as they are thought to be transformed through baryonic feedback (or because of $\Lambda$CDM shortcomings). In the case of $\Lambda$CDM collisionless simulations, the universality of the NFW profile has been tested over 20 orders of magnitude in masses of the host halos \cite{Wang:2019ftp}. While it is generally admitted that the very first halos display a steeper cusp than NFW \cite{Ishiyama:2014uoa,Colombi:2020xbv,Delos_2022}, this initial power-law rather quickly breaks down with the first mergers and always converges toward NFW. The physical origin of this attractor behaviour remains however elusive. Various attempts in the literature have discussed this question such as the mass accretion history and the role of mergers \cite{Wang:2008un,Ludlow:2013bd}, adiabatic contraction \cite{2010arXiv1010.2539D},  maximum entropy arguments in the context of violent relaxation \cite{2013MNRAS.430..121P}, mean field statistical approach \cite{Wagner:2020opz}, even numerical noise \cite{Baushev:2013ida} and, of particular relevance for our work, the initial conditions \cite{Brown:2020lxj}. In Ref.~\cite{Brown:2020lxj}, the authors showed that the universality of the NFW profile may come from a choice of primordial power spectrum in a narrow range compatible with the CMB. Most of these attempts relied on Gaussian initial conditions, see however Ref.~\cite{Colombi:2020xbv}. 

The density profiles of CDM halos formed from non-Gaussian initial statistics has been previously studied with a combination of N-body simulations and theoretical works \cite{Avila-Reese:2003cjm, Fedeli:2009mt,Smith:2010fh, Figueroa:2012ws, MoradinezhadDizgah:2013rkr, Stahl:2022did}. As changing the initial conditions changes the timing of structure formation and thus the merging history,  halos formed from different initial conditions may indeed have a different inner structure. More precisely, an excess in the positive tail of the primeval matter distribution would translate into more concentrated halos \cite{Avila-Reese:2003cjm}. Ref.~\cite{Fedeli:2009mt} used a generalized NFW profile to model halos while Ref.~\cite{Smith:2010fh} used a ``log-linear model'' to fit the ratio of the non-Gaussian to the Gaussian profile. This model has been further improved in Ref.~\cite{Figueroa:2012ws}. Finally Ref.~\cite{MoradinezhadDizgah:2013rkr}, using analytical modelling, found that for small PNG, the density profiles are not impacted by initial conditions. For larger, small scale PNG, we will show hereafter in this short paper that this result does not apply.

In Section \ref{sec:simulations}, we briefly describe our suites of simulations and how halos have been identified. In Section \ref{sec:density}, we present the measurements of the density profiles in the NFW context, and perform for the first time an analysis of their sparsity. In Section \ref{sec:abg}, we consider alternatives to the NFW profiles that better match our simulations with PNG. In Section \ref{sec:concl}, we wrap up and draw some perspectives. 

\section{Simulations' description}\label{sec:simulations}

In the following, we briefly describe our suites of simulations and halo
samples. No new simulations have been run for the present analysis, and more detailed information about those simulations can thus be found in Refs.~\cite{Stahl:2022did,Stahl:2023ccv}. 

In this paper, we will mainly study 3 $\times$ 21 collisionless (dark matter only) simulations run with the  {\sc Gadget4} code \cite{Springel:2020plp, Stahl:2022did,Stahl:2023ccv}, which we will complement in Appendix \ref{sec:appendix} with results from 3 hydrodynamical simulations run with the {\sc RAMSES} code \cite{Teyssier:2001cp}, closely following the baryonic physics of HorizonAGN \cite{Dubois:2014lxa,Dubois_2016, Stahl:2023ccv}.

We have simulated boxes of ``Gaussian'' universes as benchmarks corresponding to the standard $\Lambda$CDM models, with box-length $L$=30 Mpc/$h$. Our beyond-$\Lambda$CDM physics consists of two PNG templates with skewness: $f_{\rm NL} \sim \pm 1000$ dubbed\footnote{The sign of $f_{\rm NL}$ is opposite to the sign of the quadratic correction to a Gaussian field for the potential, hence our naming convention for the simulations.} NG$\reversedpm$. While such high values of $f_{\rm NL}$ are ruled out \cite{Planck:2019kim} at CMB scales, at scales $\mathcal{O}(10)$ Mpc/$h$, the constraints on $f_{\rm NL}$ are much looser \cite{Sabti:2020ser}. We effectively restricted the perturbative expansion around a Gaussian to such small scales since our box-length is only $L=30$ Mpc/$h$. We leave the generalization to a scale-dependent scenario, e.g. based on a fully-fledged inflationary model (eg.~\cite{Khoury:2008wj,Riotto:2010nh,Byrnes:2011gh,Jackson:2023obv,Pinol:2023oux}), for future works.

The initial conditions for our simulations were generated at $z=50$ using {\sc Monofonic} \cite{Michaux:2020yis} and we used the following cosmological parameters \cite{Stahl:2023ccv}: $\Omega_{\rm m}=0.31$, $\Omega_\Lambda=0.69$, $\Omega_{\rm b}=0.0455$, $H_0=67.7\, \rm km\,s^{-1}\, Mpc^{-1}$, $A_s=2.1 \times 10^{-9}$ and $n_{\rm s}=0.968$.
The mass resolution for the collisionless simulations is 2.6 $\times 10^7$ $M_{\odot}$ for 512$^3$ dark matter particles in the simulation box. 
We have run our simulations down to $z=0$ and saved three snapshots at $z=3$, $z=1$ and $z=0$. We identified halos using {\sc subfind} \cite{Springel:2000qu} and {\sc AdaptaHOP} \cite{Aubert:2004mu}. In order to have enough resolution to investigate the inner part of each halo, we focus in the paper on the 1000 most massive halos of each simulation. Their mass range is [$1 \times 10^{11}$ ; $2 \times 10^{14}$]~$M_\odot$. Unless otherwise mentioned, we work with virial masses $M_{200c}$ and virial radii $r_{200c}$, and we will drop the superscript $c$ for the rest of the paper. The error bars are computed using standard deviation on the mean.

\section{NFW density profiles}\label{sec:density}
 
 \subsection{Direct fit}
  \label{sec:NFWfit}

 We first aim at finding the best-fit NFW profiles describing our halos. We will later on test whether they really are a satisfactory description of the mass profiles in all cases.
 
 The universal NFW density profile discussed in Section \ref{sec:Introduction} is given by:
\begin{equation}
    \rho(r)=\frac{\rho_S}{\left(\frac{r}{r_S}\right)\left(1+\frac{r}{r_S}\right)^2},
\end{equation}
where $\rho(r)$ is the dark matter density, $\rho_S$ and $r_S$ are the scale density and radius. Instead of these two parameters, one can also work with the virial mass and concentration, or alternatively with the virial radius and concentration. The enclosed mass within a sphere of radius $r$ reads
\begin{equation}
    M(r)= 4 \pi r_S^3 \rho_S \left( \ln(1+r/r_S) - \frac{r/r_S}{1+r/r_S} \right) \, .
\end{equation}
Defining the concentration parameter $c \equiv r_{200}/r_S$, and evaluating the enclosed mass at $r_{200}$:
\begin{equation}
    \frac{\rho_S}{\rho_c}=\frac{200}{3} \frac{c^3}{\ln(1+c) - \frac{c}{1+c}},
\end{equation}
where $\rho_c$ is the critical density. We then relate the enclosed mass to the concentration $c$ and $r_{200}$:
\begin{equation}
\label{eq:enclosed mass}
    M(r)=200 \rho_c\frac{4\pi r_{200}^3}{3}\frac{\ln(1 + \frac{r c}{r_{200}}) - \frac{r c}{r_{200}(1 + \frac{r c}{r_{200}})}}{\ln(1 + c) - \frac{c}{1 + c}}.
\end{equation}
In our simulations, we first fit the logarithm of Eq.~\eqref{eq:enclosed mass} against the logarithm of the enclosed mass as measured directly from the simulations in concentric spheres. We use 30 logarithmically spaced radii, ranging from $r=$max[10 kpc/$h$ , $r_{200}/50$] up to $r=r_{200}$, closely following the prescription of Ref.~\cite{2013ApJ...762..109B}.

\begin{figure}

        \includegraphics[width=\textwidth]{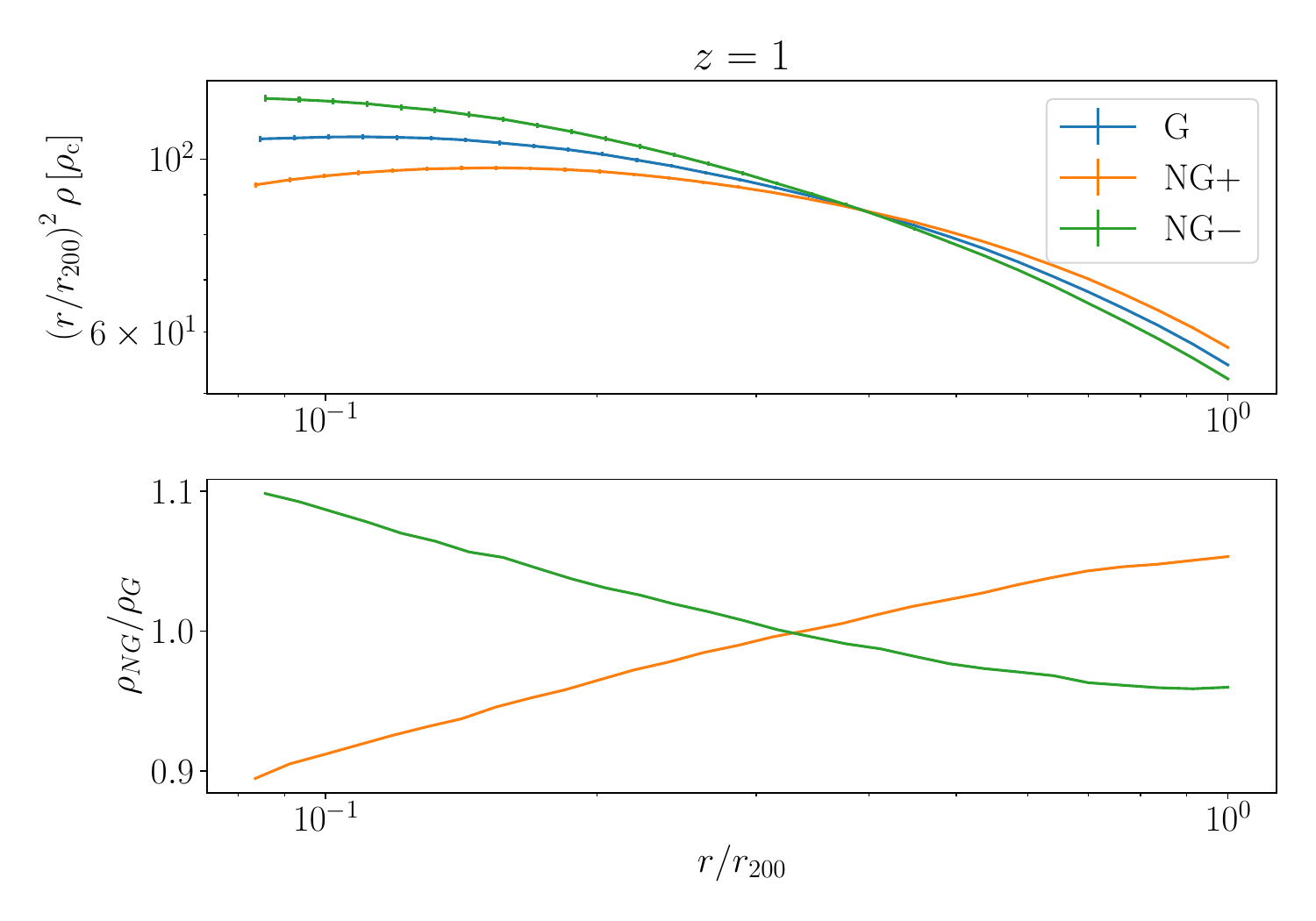}
        \caption{Average over the 21 000 halos of the density profile (not fitted to any particular parametrized profile) for the three models G, NG+ and NG- at redshift $z=1$. The error bars were calculated as the standard deviation on the mean. The NG+ halos (with negative $f_{\rm NL}$) are clearly less concentrated than their Gaussian counterparts. The opposite holds for NG- (with positive $f_{\rm NL}$).}
              \label{fig:meanPRO}
         \end{figure}

\begin{figure}

        \includegraphics[width=\textwidth]{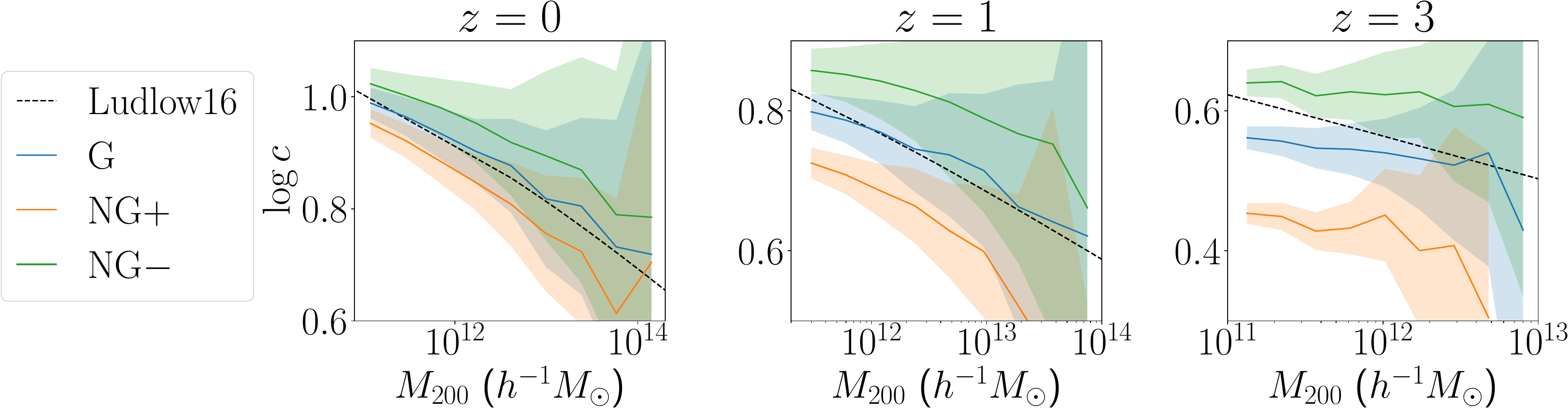}
        \caption{NFW mass-concentration relation for the three models G, NG+ and NG- at $z=0$ (left panel), $z=1$ (middle panel) and $z=3$ (right panel). The shaded regions correspond to standard deviation on the mean.}
              \label{fig:mc}
         \end{figure}

In Figure \ref{fig:meanPRO}, we display the average density profile for the models under study, without any fitting. 
It is immediately apparent that halos have different concentration in the two non-Gaussian models with positive and negative $f_{\rm NL}$. After fitting NFW profiles, we display the mass-concentration relation for our fitted halo samples in Figure \ref{fig:mc}, at three different redshifts ($z=3$, $z=1$ and $z=0$). The theoretical line from Ref.~\cite{Ludlow:2016ifl} has been plotted with the python package {\sc colossus} \cite{Diemer:2017bwl}. The fit of \cite{Ludlow:2016ifl} is in reasonable agreement with our Gaussian halos, while NG+ (negative $f_{\rm NL}$) is systematically less concentrated and NG- (positive $f_{\rm NL}$) is more concentrated. This confirms well known results from, e.g., Refs.~\cite{Avila-Reese:2003cjm, Smith:2010fh, Figueroa:2012ws}.

 \subsection{Using sparsity to evaluate the quality of the NFW description}
 The calculation of the concentration statistics depicted in Subsection \ref{sec:NFWfit} relied on a fit to NFW profiles. However, it had nothing to say on whether the NFW profile is indeed a correct description of the halo density profiles in the non-Gaussian cases under consideration, which is a more profound question than the one relating to the mass-concentration relation.

The halo sparsity, namely the ratio between cumulative masses enclosing two different overdensities, provides a non-parametric estimate of the halo mass distribution that encodes information beyond the simple NFW description \cite{Balmes:2013hqa, Corasaniti:2017yts,Corasaniti_2019,Corasaniti_2020,Corasaniti_2022,Richardson:2021yvu,Richardson:2022cvd}. It is defined as:
 \begin{equation}
 \label{eq:DEFspar}
     s_{\Delta}\equiv\frac{M_{200}}{M_{\Delta}},
 \end{equation}
the ratio of the masses enclosed in spheres of radius $r_{200}$ and $r_{\Delta}$. For all $\Delta>0, r_{\Delta}$ is the radius within which the average density is equal to $\Delta$ times the critical density.

Typical values for $\Delta$ include 2500, 1000 and 500, with $r_{200}>r_{500}>r_{1000}>r_{2500}$. 
This means that the sparsities give an image of the inner structure of the halo without relying on any parametric density function. It has been shown to contain a wealth of cosmological \cite[e.g.,][]{Corasaniti:2017yts} and astrophysical \cite[e.g.,][]{Richardson:2021yvu,Richardson:2022cvd} information. Regarding observational prospects, it is also simpler to measure than a full profile (see Ref.~\cite{Tamara} for a full presentation of the sparsity).

For NFW profiles, using Equation \eqref{eq:enclosed mass} allows one to exhibit a one to one correspondence between the sparsity and the NFW concentration:
\begin{equation}
\label{eq:StoC}
   s_{\Delta}^{{\rm NFW}}= \frac{\ln(1 + c) - \frac{c}{1 + c}}{\ln(1 + \frac{r_{\Delta} c}{r_{200}}) - \frac{r_{\Delta} c}{r_{200}(1 + \frac{r_{\Delta} c}{r_{200}})}}.
\end{equation}

\begin{figure}

        \includegraphics[width=.75\textwidth]{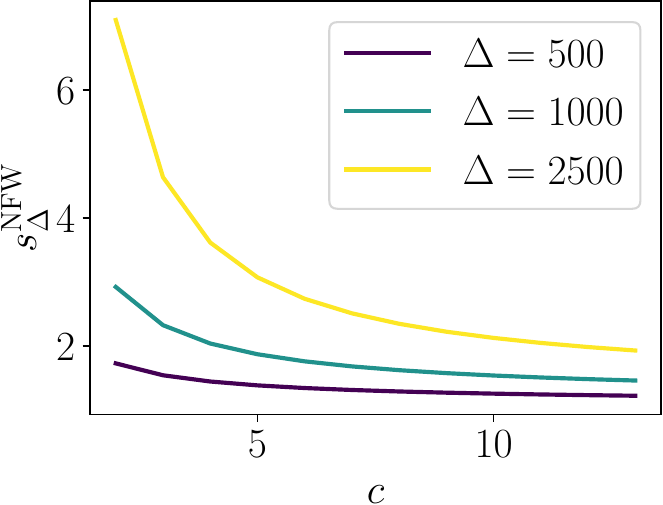}
                \caption{Three typical sparsities as a function of the concentration of the NFW profile obtained using Eq.~\eqref{eq:StoC}.}
                      \label{fig:3.2tamara}
         \end{figure}

A graphical representation can be obtained in Figure 3.2 of Ref.~\cite{Tamara}, that we reproduce in Fig.~\ref{fig:3.2tamara}.

\begin{figure}

        \includegraphics[width=\textwidth]{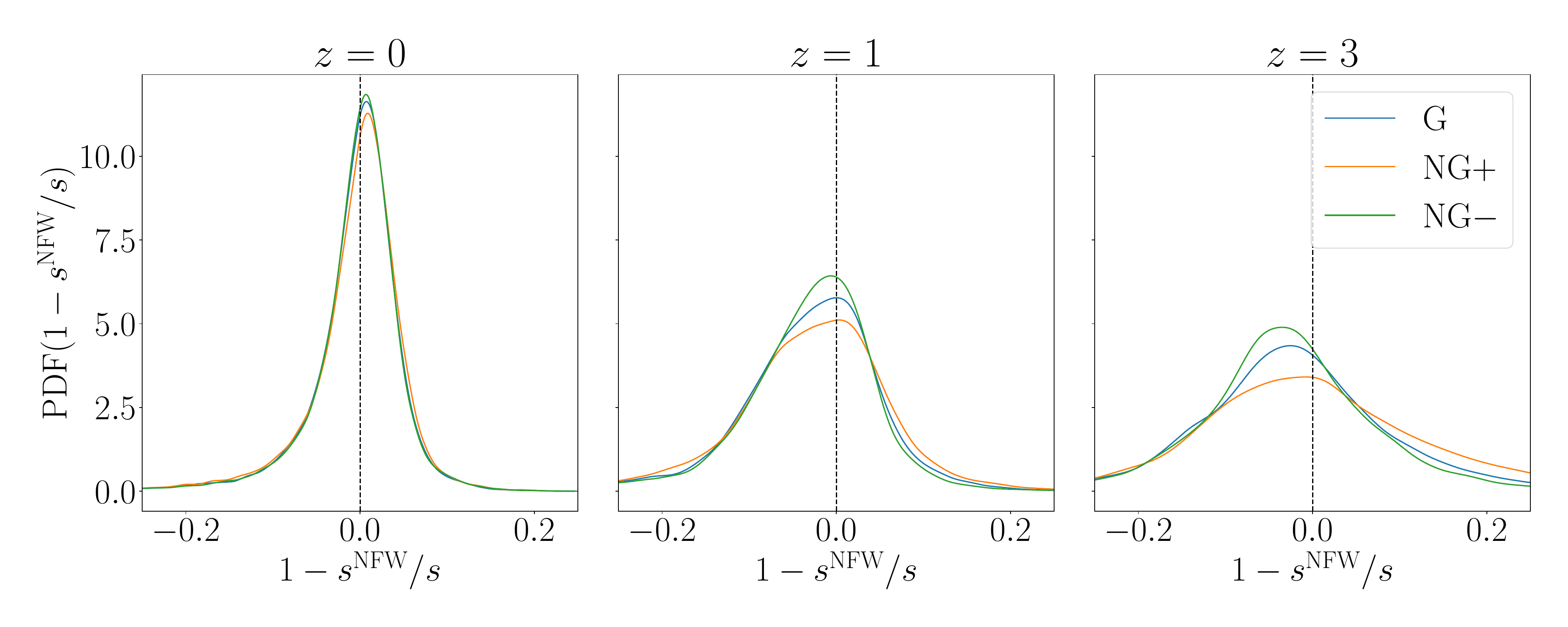}
                \caption{Distributions of the relative difference between the sparsities $s\equiv s_{500}$ directly measured in our Gaussian and non-Gaussian simulations and the sparsities $s^{\rm NFW}\equiv s^{\rm NFW}_{500}$ inferred from a NFW fit as per Eq.~\eqref{eq:StoC}. If the quantity plotted is positive/negative for a given halo, it means that the halo is sparser/less sparse : namely, the halo has less/more mass at $r_{500}$ that what is predicted by the NFW fit. Owing to its lower PDF at the mode and to its larger tails, it appears that NG+ is less close to NFW than G and NG-.}
                      \label{fig:Ratios}
         \end{figure}

  In Figure \ref{fig:Ratios}, we compare the sparsity $s\equiv s_{500}$ measured directly from the enclosed mass to the sparsity $s^{\rm NFW}$ deduced from the NFW fit of Section \ref{sec:NFWfit}, with Eq.~\eqref{eq:StoC}.
To have a deeper understanding of our halo sample, we do not only focus on the median of the halo but plot the full probability density function of the relative ratios of the sparsities as in Figure 1 of Ref.~\cite{Richardson:2021yvu}.
At $z=0$, there is no obvious deviation from NFW compared to the Gaussian case, but the tails of the NG+ simulation indicate a possible deviation at high redshift. In order to examine the merit of this impression, we then perform a second test, not relying on the fitting procedure of Section \ref{sec:NFWfit}:
\begin{enumerate}
    \item We measure $s_{500}$ from the simulation using Eq.~\eqref{eq:DEFspar}.
    \item We use twice Eq.~\eqref{eq:StoC} to obtain $c$ and then $s_{2500}$.
    \item We then measure $s_{2500}$ from the simulation.
    \item $\Delta s_{2500}$ represents the difference between the two values found in 2. and 3.
\end{enumerate}

\begin{figure}

        \includegraphics[width=\textwidth]{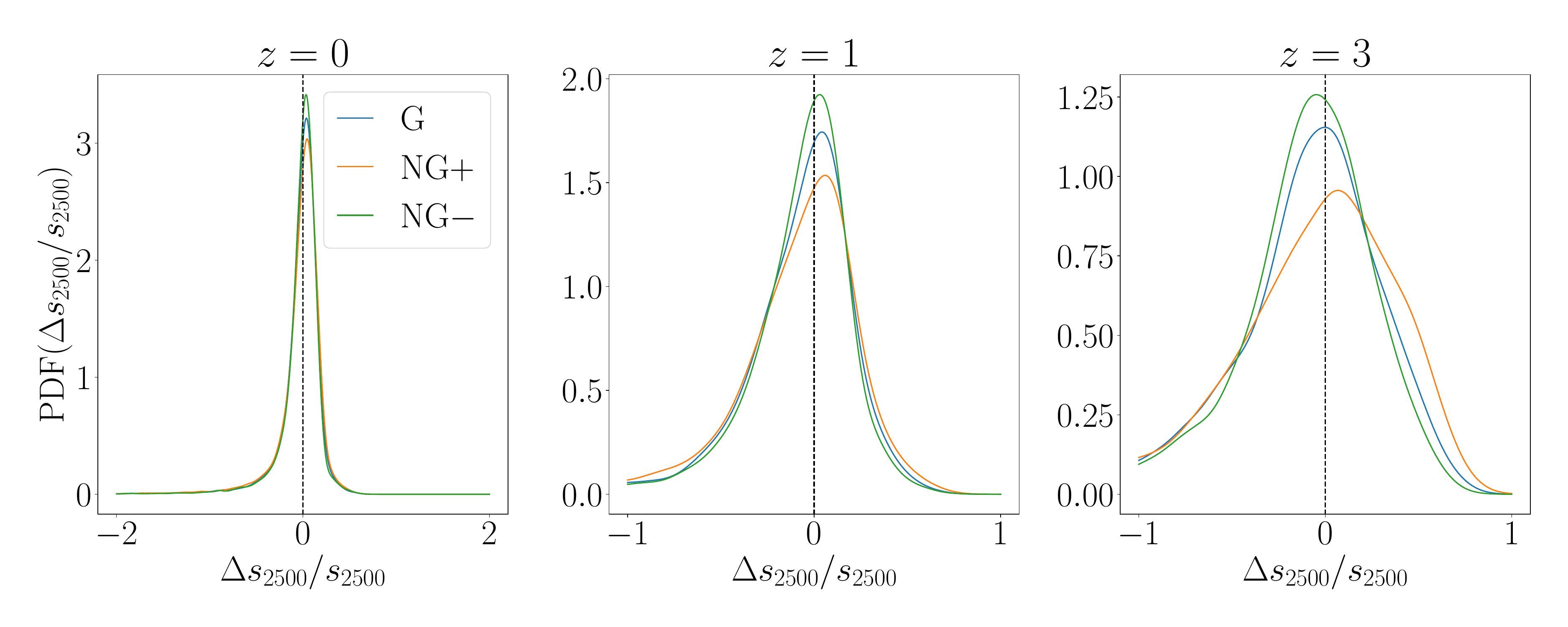}
                \caption{Distributions of the relative ratios of sparsities at $r_{2500}$: one directly measured and the second inferred from a measure at $r_{500}$ assuming the NFW relation of Eq.~\eqref{eq:StoC}. See points 1-4 after Eq.~\eqref{eq:StoC} for more details. Note that no fit to a family of profile were involved to obtain this curve, though the results are in agreement with Fig.~\ref{fig:Ratios} which relied to a fit to NFW. If the quantity plotted is positive/negative for a given halo, it means that the halo is sparser/less sparse than expected from NFW. Again, the lower PDF at the mode and larger tails of NG+ indicate that it is less well described by a NFW profile than the others.}
                      \label{fig:TB}
         \end{figure}

 In Figure \ref{fig:TB}, we compare the distribution of $\Delta s_{2500}/s_{2500}$, where $s_{2500}$ corresponds to the sparsity directly measured as in the aforementioned point 3.
This second test conclusively confirms the deviation of NG+ from NFW at intermediate and high redshift. Moreover, as could already be hinted from Fig.~\ref{fig:Ratios}, the NG- simulation is even {\it closer} to NFW than the Gaussian case as it has a more peaked distribution around 0.

This motivates us to relax the hypothesis of a NFW halo and explore whether it is possible to find a more general family of profiles, in the spirit of \cite{Smith:2010fh, Figueroa:2012ws}, that adjust better the halos in simulations with PNG, especially for the NG+ (negative $f_{\rm NL}$) data set.      

\section{Other profiles for non-zero $f_{\rm NL}$}
\label{sec:abg}
\subsection{Minimal extension to NFW}
A large number of profiles able to describe the density distribution of spherical dark matter halos exist in the literature~\cite[e.g.,][]{Einasto,Jaffe,Hernquist,Dehnen,Evans,Tremaine,Burkert,Merritt,2010arXiv1005.0411C, Read, Lazar,Jo}.
Among these, a very general family of profiles extending the NFW profile is the following parametrization \cite{Zhao}: 
\begin{equation}
    \rho(r)=\frac{\rho_S}{(\frac{r}{r_S})^{\gamma} \left[1+\left(\frac{r}{r_S}\right)^{\alpha}\right]^{\frac{\beta-\gamma}{\alpha}}},
\end{equation}
where $\gamma$ represents the (logarithmic) inner slope, $\beta$ represents the outer slope, and $\alpha$ controls the transition between the two regions. The NFW profile is recovered for ($\alpha$,$\beta$,$\gamma$)  = (1,3,1). 

The enclosed mass function for this profile reads:
\begin{equation}
\label{eq:abg}
\begin{split}
    M(r)& =  4 \pi r_S^3 \int_0^{r} d r' r'^2 \frac{\rho_S}{r'^{\gamma}(1+r'^{\alpha})^{\frac{\beta - \gamma}{\alpha}}} \\
    & = 4 \pi r_S^3 \rho_S \left(\frac{r}{r_S} \right)^{3-\gamma} \frac{{}^2F_{1}\left(\frac{3-\gamma}{\alpha},\frac{\beta-\gamma}{\alpha},\frac{3+\alpha-\gamma}{\alpha},-(r/r_S)^\alpha\right)}{3-\gamma} \\
    & \equiv 4 \pi r_S^3 \rho_S G_{\alpha \beta \gamma}(r/r_S),
    \end{split}
\end{equation}
where ${}^2F_{1}$ is the hypergeometric function, see eg.~Ref.~\cite{Begue:2017lcw} for a definition and an application. In the NFW limit, $G_{1 3 1}(r/r_S)= \ln(1+r/r_S) - \frac{r/r_S}{1+r/r_S} $. 

Still defining $c \equiv r_{200}/r_S$, the enclosed mass can be rewritten as
\begin{equation}
    \label{eq:enclosed massABG}
    M(r)=200 \rho_c\frac{4\pi r_{200}^3}{3}\frac{G_{\alpha \beta \gamma}(rc/r_{200})}{G_{\alpha \beta \gamma}(c)}.
\end{equation}
In our case, $r_{200}$ can be measured directly from the simulation, hence this reduces the profile to 4 parameters. However, as we only have a one-dimensional profile to fit, and as it is clear from the sparsity analysis that only a minimal extension to NFW is needed,  we will fix some of the parameters to break the degeneracies. 

In Ref.~\cite{Smith:2010fh} (see also Ref.~\cite{Figueroa:2012ws}), the ratios of the NG profiles to the Gaussian ones have been described by a log linear two-parameters model:
\begin{equation}
\frac{\rho_{\rm {NG}}}{\rho_{\rm {G}}} = 1 + m \log(r/r_X),
\label{eq:Smith10}
\end{equation}
with a length-scale $r_X$ and a shape transformation $m$.
This model was found to encapsulate the corrections due to a positive or negative $f_{\rm NL}$, that were qualitatively equivalent to what we found in Figure 8 of Ref.~\cite{Stahl:2022did}. As a first exercise, we fitted with Eq.~\eqref{eq:Smith10} the ratios of our profiles and have found a qualitative agreement with Ref.~\cite{Smith:2010fh}.

\begin{figure}

        \includegraphics[width=\textwidth]{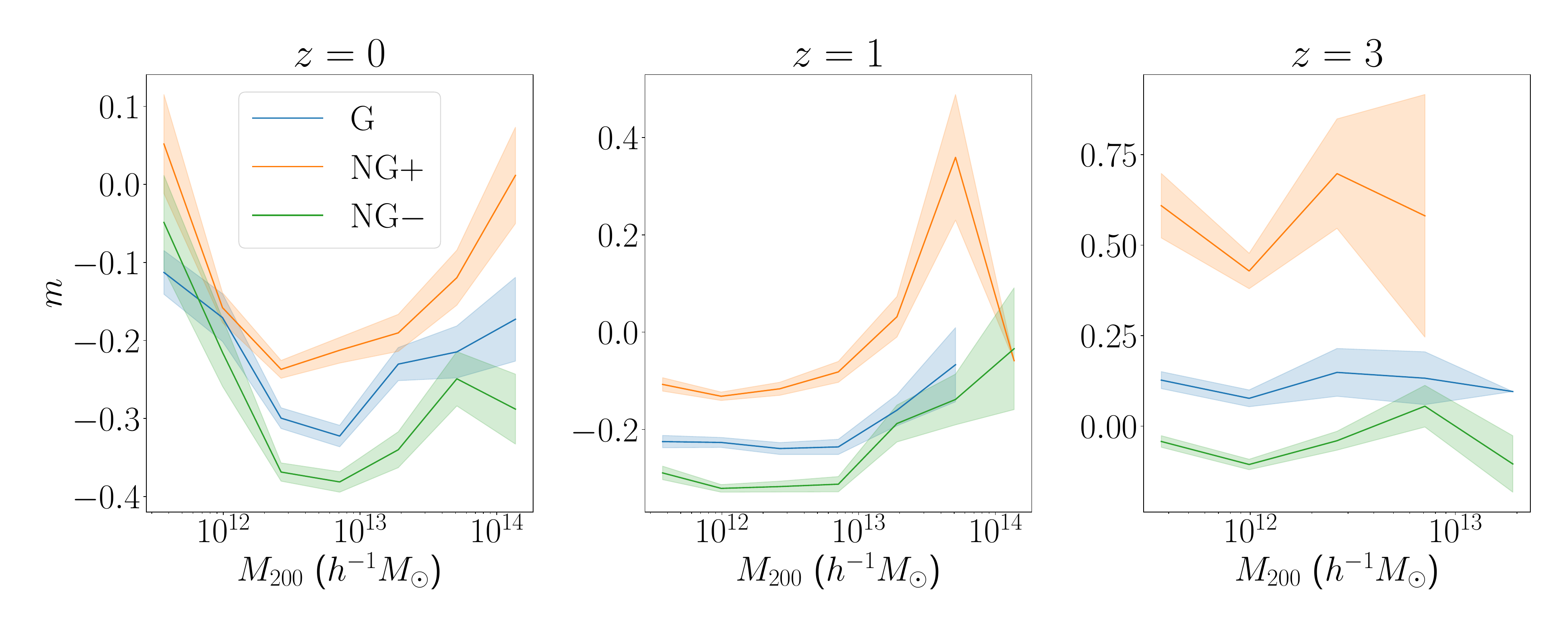}
                \caption{Resulting correction for the inner slope of the NFW profile, as a function of the mass of the halo. $m \equiv 1 - \gamma$, see Eq.~\eqref{eq:abg}. The uncertainties on mean $m$ are sometimes zero at high mass when only one halo is present in the mass bin.}
                      \label{fig:mM}
         \end{figure}

Then, motivated by Eq.~\eqref{eq:Smith10}, which for small $m$ can be rewritten $\left(\frac{r}{r_X} \right)^m$, we propose to describe the halos in simulations with PNG with a minimal extension to NFW, given by Eq.~\eqref{eq:abg} with $(\alpha, \beta, \gamma)=(1,3,1-m)$: they present the advantage of leaving the outer slope equal to three but allowing the inner asymptotic slope to vary. Note also that $r_X$ and $\rho_S$ are totally degenerate and therefore determined from $M_{200}$ and $r_{200}$.
Following the same procedure discussed in Section \ref{sec:density}, we fit the halos with this minimal extension of NFW profiles. To have enough resolution to avoid numerical artefacts, we now restrict to halos with mass larger than $3 \times 10^{11} {\rm M}_{\odot}$. We display the fitted values of $m$ in Figure \ref{fig:mM}. 
Halos naturally tend to prefer a slightly more cuspy central slope than NFW at $z=0$, but as we shall see below, the improvement in the fit is marginal in the Gaussian and NG- cases.

At all three redshifts considered, it appears clearly that the NG+ simulations prefer higher values of $m$ than G and NG-, namely shallower inner slopes, especially so at high redshifts, in qualitative agreement with our findings based on sparsity. The NG- simulations, on the other hand, appear to be a bit more cuspy than the Gaussian case. 

In summary, we have qualitatively found, both from the analysis of the sparsity in Section \ref{sec:density} and from fits with a minimal extension to NFW, that halos in the NG+ simulations seem to be at odds with the expectations from a universal NFW profile, preferring shallower inner slopes. 

We now quantify this by checking if something has been gained with our fits based on a minimal extension to the NFW profile. For this, we consider the Akaike Information Criterion (AIC) suited for nested fitting situations such as the one considered here: $m=0$ recovers the NFW profile. This criterion penalizes the goodness of the fit by taking into account the number of degrees of freedom used to perform the fit.

\begin{figure}

        \includegraphics[width=\textwidth]{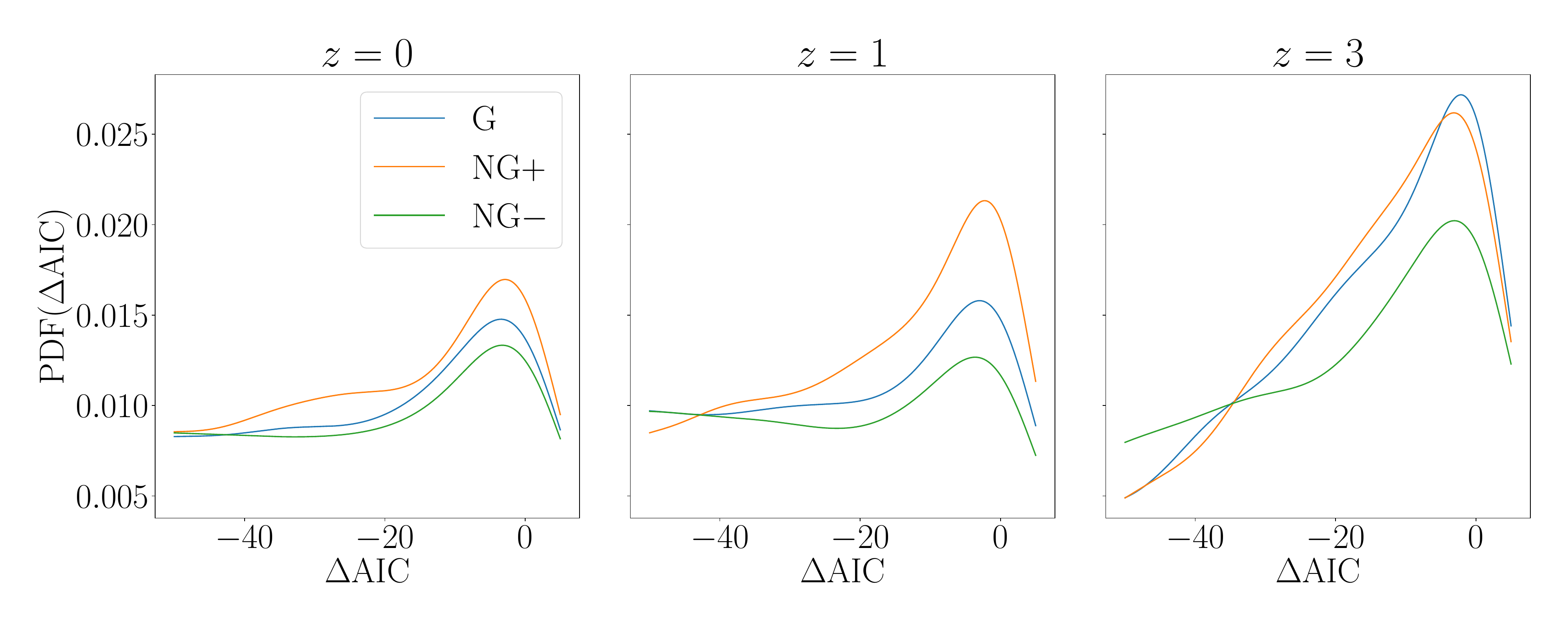}
                \caption{Distribution of the difference of the Akaike Information Criterion fitted by our minimal extension of NFW with that obtained with a NFW profile, for the Gaussian and non-Gaussian models. A negative value indicates that the minimally extended NFW is a better adjustment to the density profiles than NFW. Especially at $z=1$, NG+ improves the most compared to its NFW value.}
                      \label{fig:dAIC}
         \end{figure}

We compute the difference $\Delta$AIC between the minimal NFW extension's AIC and the NFW's AIC, defined as $\Delta$AIC. In order to check if something has been gained, we also restrict to halos with $|\Delta {\rm AIC}|<50$. Doing so allows to keep halos which improve at maximum 3 times the mean improvement in AIC of the Gaussian case. Indeed halos in the process of merging, or generally out of equilibrium, could lead to an improvement even in the Gaussian case, but are of no interest for our present investigation. We display the resulting PDF in Figure \ref{fig:dAIC}.

The minimally extended NFW model generally improves the fit of NG+ compared to the NFW model\footnote{Some larger improvements are obtained, at larger $|\Delta$AIC|, in the Gaussian case: these halos are typically not at equilibrium.} with a lot of halos with $\Delta$AIC down to $-40$. Interestingly, this is most visible at $z=1$. We therefore conclude, from our sparsity analysis combined with our present parametric analysis, that $z=1$ halos carry the most interesting information about PNG. 

\subsection{Einasto profile}
\begin{figure}

        \includegraphics[width=\textwidth]{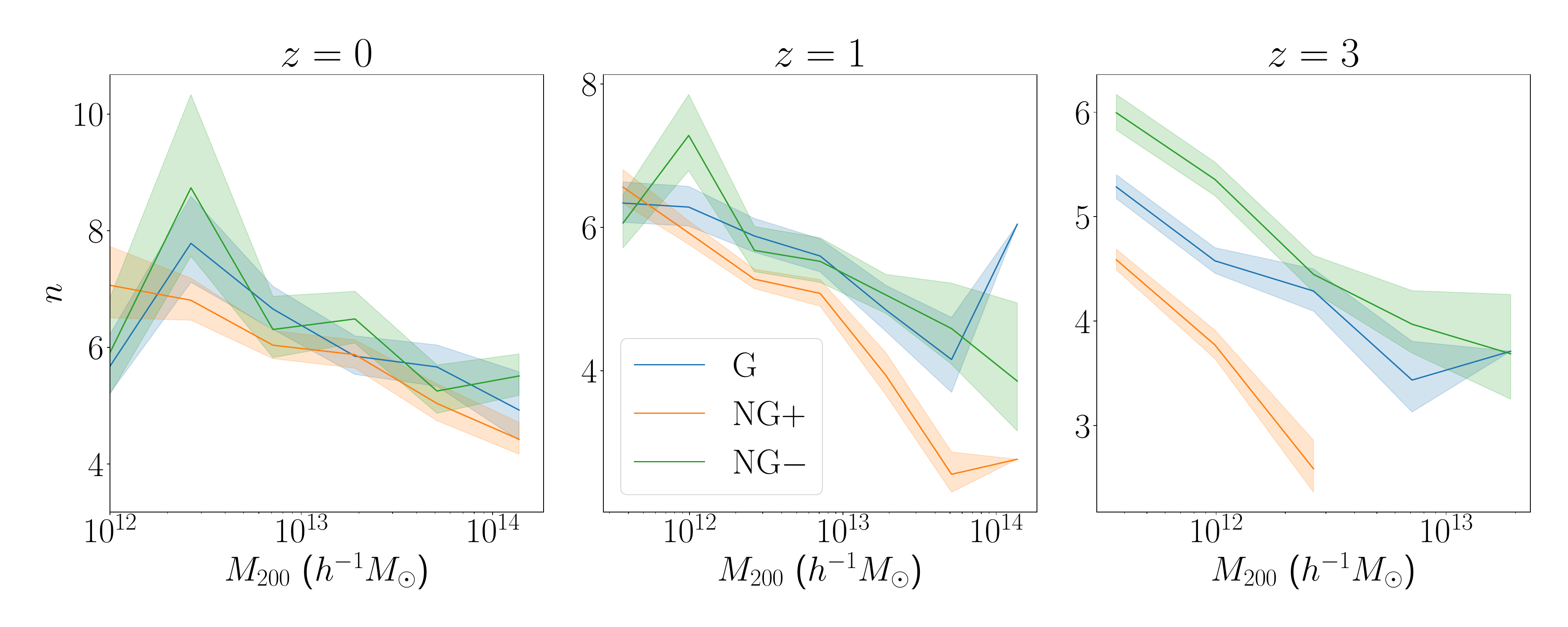}
                \caption{Value of the parameter $n$ of the Einasto profile \eqref{eq:Einasto} fits as a function of the mass of the halo. }
                      \label{fig:Einasto}
         \end{figure}

The analysis of the previous subsection focused on a minimal extension, which only changed the inner slope of the NFW profile. The NFW profile transitions gradually from $\rho \propto r^{-3}$ at large radii to $r^{-1}$ at small radii, whereas the minimal extension considered in the previous subsection transitions to $r^{-(1-m)}$ , where $m$ is free. However, the simulations do not resolve the inner cusp, especially at low masses, hence the values of $m$ inferred from the previous fits are essentially indicating a change of slope compared to NFW in the transition region, rather than probing the inner cusp.
To confirm this, we also consider as an alternative parametrization the Einasto profile \cite{Einasto,Ghari}:
\begin{equation}
\label{eq:Einasto}
    \rho(r)=\rho_{-2} \exp \left[-2n \left( \left(\frac{r}{r_{-2}} \right)^{1/n} -1 \right) \right],
\end{equation}
where $r_{-2}$ is the radius at which the density has a slope of -2, $\rho_{-2}$ is the density at that radius and $n$ describes how sharply the profile steepens at low $r$. The meaning of the index $n$ is closely related to the parameter $\alpha$ in our Eq.~\eqref{eq:abg}. 
The Einasto profile can describe the subtle variations with respect to NFW for the CDM halo densities in the Gaussian case. Its parameters vary with the mass of the halo, the redshift \cite{Gao:2007gh,Dutton:2014xda} and the initial power spectrum \cite{Ludlow:2016qow}. We performed a fit on the density profiles themselves for the three parameters ($n$, $\rho_{-2}$, $r_ {-2}$). The results for the index $n$ are displayed on Figure \ref{fig:Einasto}. We recover the expected dependence of $n$ on the halo mass for the Gaussian case, and we qualitatively confirm the results of the previous subsection, i.e.~that the NG+ simulations generically transition to a shallower central density (not well resolved in our simulations) with a lower value of $n$ than in the Gaussian case.

\begin{figure}
        \includegraphics[width=\textwidth]{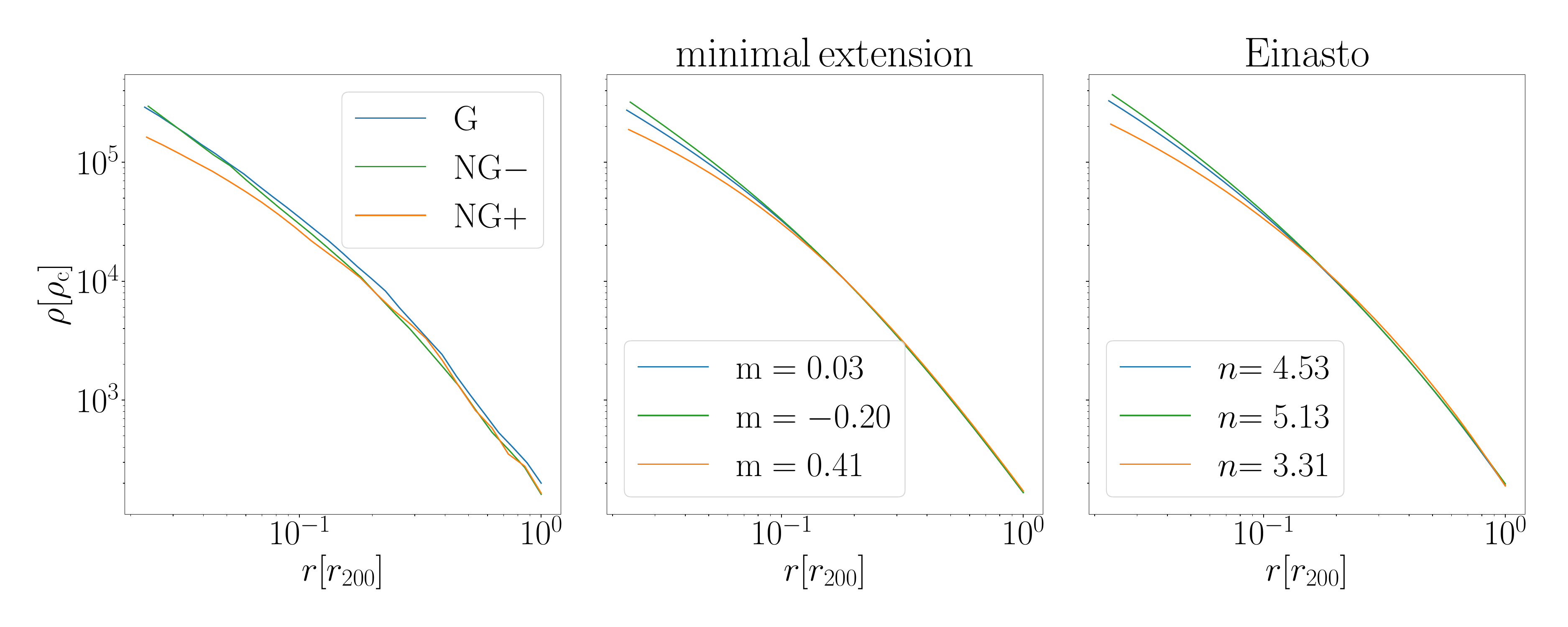}
                \caption{At $z=0$, the density profile of the most massive halo of one simulation (left panel), the fit to the minimal extension of NFW (middle panel) and the Einasto fit (right panel).}
                      \label{fig:MMprofile}
         \end{figure}

Finally, in Figure \ref{fig:MMprofile}, to illustrate these results, we select the most massive halo of one simulation and show its density profile, along with the fit to the minimal extension to NFW and to Einasto, illustrating this shallower transition for the NG+ case.

\section{Conclusions and perspectives}
\label{sec:concl}
In this short paper, we studied in depth the impact of small ($\sim 30$ Mpc) scale PNG on the internal structure of dark matter halos formed in a suite of collisionless cosmological simulations with $f_{\rm NL}={-1000, 0, 1000}$ (dubbed NG+, G, NG-) at three different redshifts $z={3, 1, 0}$. Since such PNG are known to change the timing of structure formation and the merger trees, they can potentially affect the inner structure of halos. In particular, knowing whether the NFW profile remains an attractor solution for the halo density profiles in the presence of PNG is an important question, which can, in turn, shed light on the reasons behind the universality of this profile in the $\Lambda$CDM context. 

We start by recovering a series of already known results, namely that halos formed in simulations with negative $f_{\rm NL}$ (NG+) are less concentrated than in the Gaussian case, while the opposite is true in the case of a positive $f_{\rm NL}$. Then, for the first time, we perform a detailed analysis of the sparsity of the halos formed in the different simulations. The distribution of sparsities indicates qualitatively that the NFW profile remains a relatively good description at $z=0$, even though the NG+ distribution tends to be more skewed, indicating that the NFW description might not be sufficient. This becomes more marked at $z=1$ and $z=3$. To confirm this qualitative finding based on sparisties, we then adjust the profiles with a minimal extension to NFW by varying its asymptotic inner slope. While some halos that are out of equilibrium can be better described by such a minimal extension than by NFW (based on the Akaike Information Criterion) even in the Gaussian case, restricting to $|\Delta$AIC| up to 50, it is clear that the minimal extension leads to a significant improvement in the NG+ case at $z=0$ and $z=1$. Combined with the sparsity analysis and a fit to the Einasto profile, we conclude that the $z=1$ halos are those carrying the most interesting information in cosmologies with negative $f_{\rm NL}$ (NG+). Since this internal structure of halos is set by the initial conditions, PNG typically lead to a different redshift dependence than other extensions to $\Lambda$CDM, which allows them to leave unique signatures. The same should be true for signatures of feedback. We provide a preliminary analysis at $z=0$ of hydrodynamical simulations in the Appendix, but exploring a larger suite of simulations at higher redshifts will be needed to explore the degeneracies (and non-degeneracies) with feedback. 

In the future, we will need to extend our analysis to a large suite of hydrodynamical simulations at various redshifts, as well as consider larger boxes with scale-dependent PNG and/or with the full PDF. We will explore whether Machine Learning methods lead to the same result concerning the non-universality of NFW in the presence of PNG \cite{Lucie-Smith:2022uvv,Lucie-Smith:2023kue,WT1, WT2}. Since we have mostly concentrated here on the inner structure of halos, it will also be interesting to consider the effect on the outer slope, splashback radius, and large-scale triaxility of the halos.

\acknowledgments
BF, RI, and CS acknowledge funding from the European Research Council (ERC) under the European Union's Horizon 2020 research and innovation program (grant agreement No.\ 834148). CS, BF and RI thank Wassim Tenachi for stimulating discussions. This work has made use of the Infinity Cluster hosted by the Institut d'Astrophysique de Paris.

\section*{Softwares}
The analysis was partially made using  \href{https://yt-project.org/}{YT} \cite{Turk:2010ah}, as well as IPython \cite{Perez:2007emg}, Matplotlib \cite{Hunter:2007ouj}, NumPy \cite{vanderWalt:2011bqk} and SciPy \cite{Virtanen:2019joe}, especially curve fit and \href{https://github.com/lmfit/lmfit-py}{lmfit}.

\section*{Authors' Contribution}
In his master thesis, NM carried out a preliminary analysis similar to section 3.1 in consultation with CS. Subsequently CS performed the more systematic study described in this work in consultation with BF. CS and BF drafted the manuscript. All the authors improved it by their comments.

\section*{Carbon Footprint}
In this work, we re-used existing simulation data. That severely limits its carbon footprint.
Following Ref.~\cite{berthoud} to convert\footnote{Including the global utilisation of the cluster and the pollution due to the electrical source, the conversion factor is 4.7 gCO2e/h core} the number of CPU hours required to postprocess the simulations in order to obtain the data for this work, we have used 100 kgCO2eq.

\appendix
\section{Hydrodynamical results}
\label{sec:appendix}
\begin{figure}

        \includegraphics[width=\textwidth]{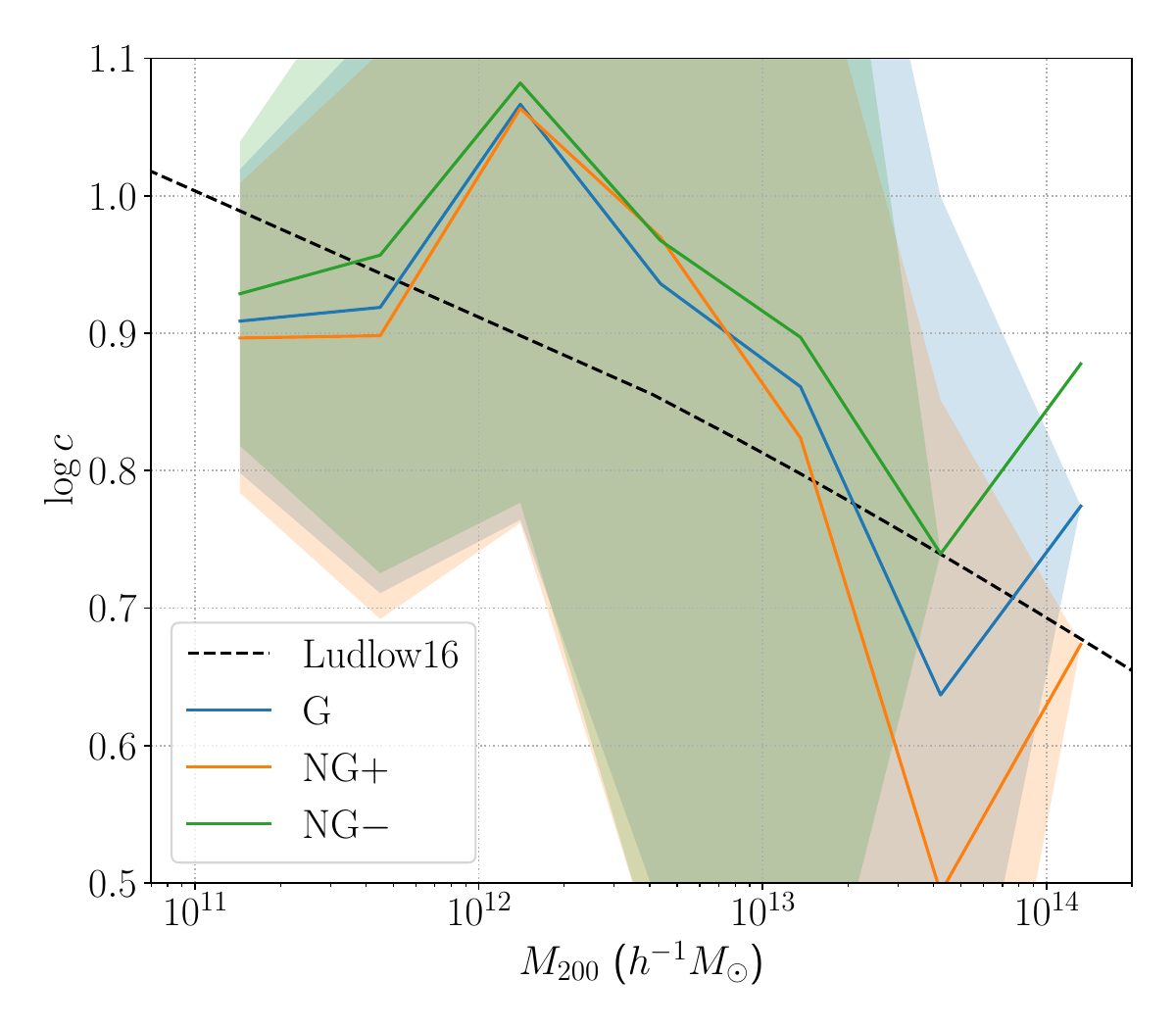}
        \caption{Mass-concentration relation for the three models G, NG+ and NG-. The uncertainties are sometimes zero at high mass when only one halo is present in the mass bin.}
              \label{fig:mcH}
         \end{figure}

In this Appendix, we complement the study of the 3 $\times$ 21 dark matter only simulations with the most salient corresponding results for the dark matter particles present in our hydrodynamical simulations, with baryonic sub-grid physics included. The resolution of the dark matter particles in these hydrodynamical simulations is 2.2 $\times 10^7$ $M_{\odot}$. In that case, we have only 1000 halos, so the statistics is poorer but the trends are similar. We only show the results at $z=0$.
In Figs.~\ref{fig:mcH}, we present the mass-concentration relation. As in Figure \ref{fig:mc}, the NG+ model is less concentrated than the G model, itself less concentrated than the NG- model. 

\begin{figure}

        \includegraphics[width=\textwidth]{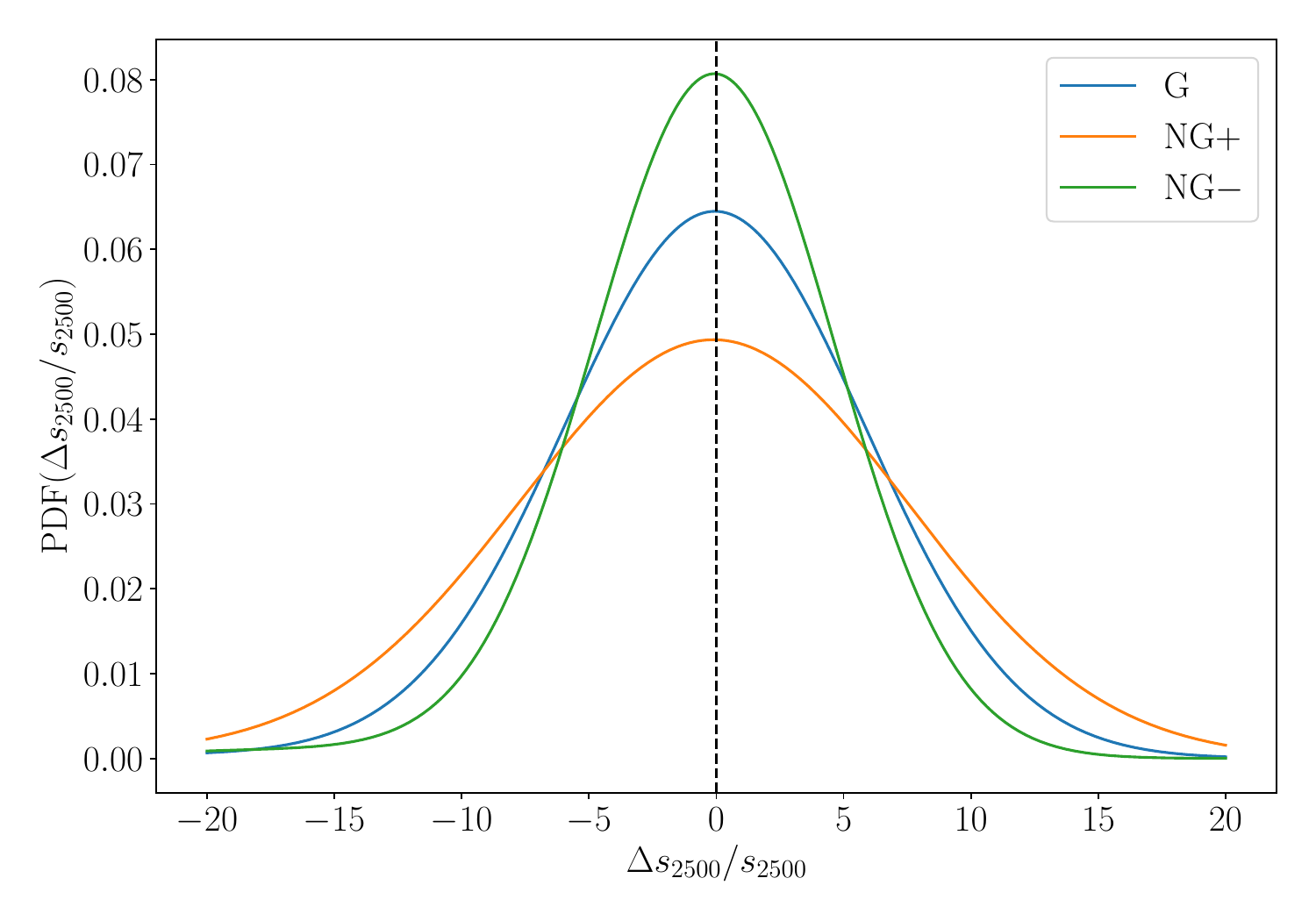}
                \caption{Distribution of the relative ratios of sparsities at $r_{2500}$: one directly measured and the second inferred from a measure at $r_{500}$. See points 1-4 after Eq.~\eqref{eq:StoC} for more details. This test is a null test in the sense that we are computing the same quantity with two methods that should agree if the halos are well described by a NFW profile. We find that NG+ is the less close to a NFW profile.}
                      \label{fig:TBH}
         \end{figure}

 In Figure \ref{fig:TBH}, we redo the test that we proposed in points 1-4 after Eq.~\eqref{eq:StoC}. Recall that this test did \textit{not} rely on a fitting procedure, but just on enclosed mass measurements.

We confirm the results found in the main text, NG+ has more difficulty to account for a NFW profile, as it presents larger tails.

\begin{figure}

        \includegraphics[width=\textwidth]{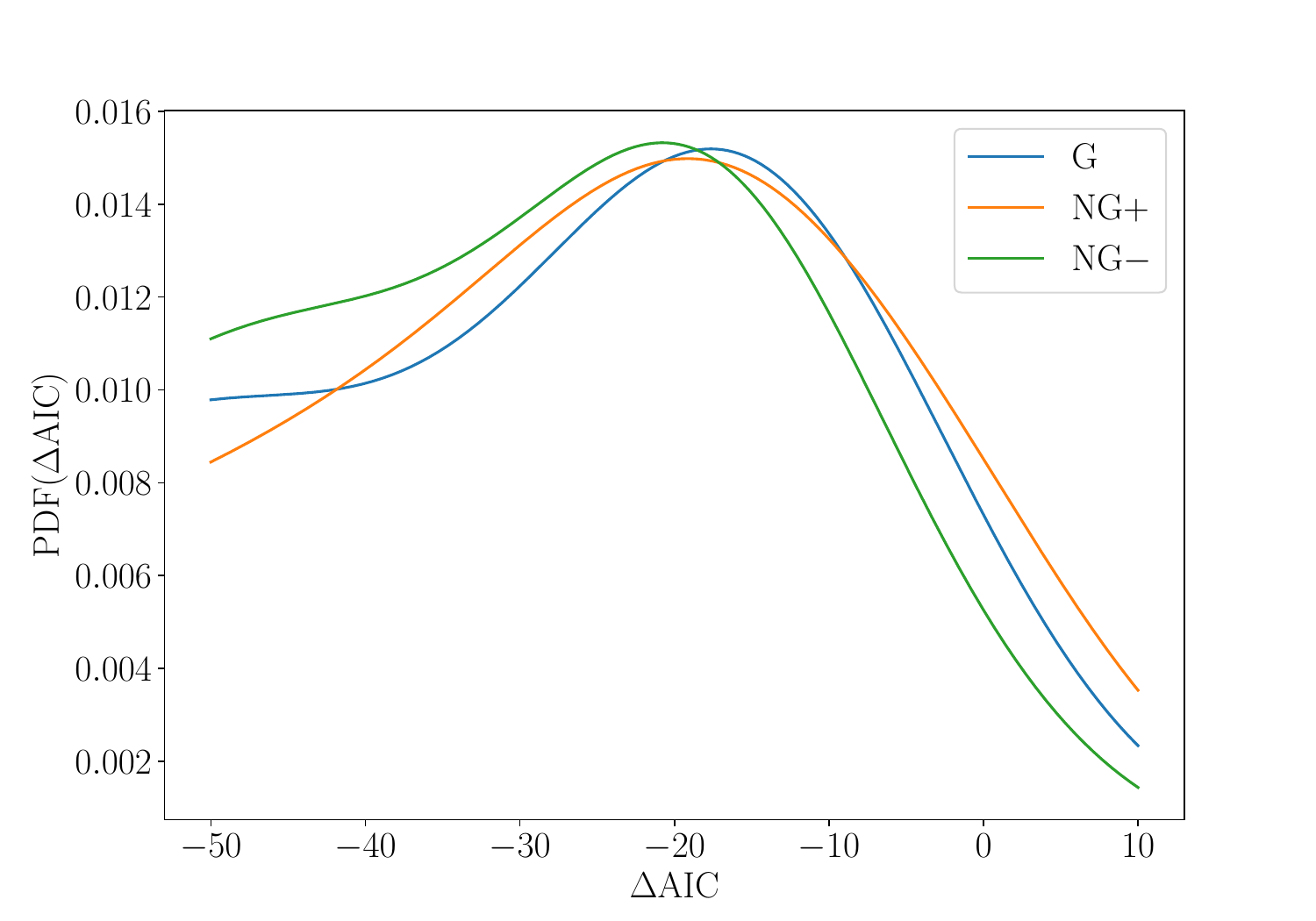}
                \caption{Distribution of the difference of the Akaike Information Criterion for the Gaussian and non-Gaussian models fitted by a minimally extended NFW and NFW profile. A negative value indicates that the minimally extended NFW is a better fit to the density profiles than NFW.}
                      \label{fig:HdAIC}
         \end{figure}

Finally, following the main text, we let the inner slope of the density profile vary. The mean values of the inner slope becomes in that case $m \sim -1$ that is $\gamma \sim 2$ (instead of $\gamma =1$ for NFW), because of the adiabatic contraction of the halos. No clear difference could however be exhibited between the models with the statistics available to us: in Fig.~\ref{fig:HdAIC}, we display the improvement of the AIC. Following the main text, we present only halos mildly improved by the minimally extended NFW, that is $|\Delta$AIC|<50. While every model gains by opening the inner slope of the fit, NG+ does not improve over the Gaussian model in that case.

\bibliographystyle{JHEP.bst}
\bibliography{ref.bib}

\providecommand{\href}[2]{#2}\begingroup\raggedright\begin{thebibliography}{100}

\bibitem{Abdalla}
E.~{Abdalla}, G.~F. {Abell{\'a}n}, A.~{Aboubrahim}, A.~{Agnello},
  {\"O}.~{Akarsu}, Y.~{Akrami} et~al., \emph{{Cosmology intertwined: A review
  of the particle physics, astrophysics, and cosmology associated with the
  cosmological tensions and anomalies}},
  \href{http://dx.doi.org/10.1016/j.jheap.2022.04.002}{\emph{Journal of High
  Energy Astrophysics} {\bf 34} (June, 2022) 49--211},
  [\href{https://arxiv.org/abs/2203.06142}{{\tt 2203.06142}}].

\bibitem{ElGordo}
R.~J. {Foley}, K.~{Andersson}, G.~{Bazin}, T.~{de Haan}, J.~{Ruel}, P.~A.~R.
  {Ade} et~al., \emph{{Discovery and Cosmological Implications of SPT-CL
  J2106-5844, the Most Massive Known Cluster at z>1}},
  \href{http://dx.doi.org/10.1088/0004-637X/731/2/86}{\emph{\apj} {\bf 731}
  (Apr., 2011) 86}, [\href{https://arxiv.org/abs/1101.1286}{{\tt 1101.1286}}].

\bibitem{Finkelstein:2013lfa}
S.~L. Finkelstein et~al., \emph{{A Rapidly Star-forming Galaxy 700 Million
  Years After the Big Bang at z=7.51}},
  \href{http://dx.doi.org/10.1038/nature12657}{\emph{Nature} {\bf 502} (2013)
  524}, [\href{https://arxiv.org/abs/1310.6031}{{\tt 1310.6031}}].

\bibitem{Labbe}
I.~{Labb{\'e}}, P.~{van Dokkum}, E.~{Nelson}, R.~{Bezanson}, K.~A. {Suess},
  J.~{Leja} et~al., \emph{{A population of red candidate massive galaxies 600
  Myr after the Big Bang}},
  \href{http://dx.doi.org/10.1038/s41586-023-05786-2}{\emph{\nat} {\bf 616}
  (Apr., 2023) 266--269}, [\href{https://arxiv.org/abs/2207.12446}{{\tt
  2207.12446}}].

\bibitem{MBK2}
M.~Boylan-Kolchin, \emph{{Stress testing \ensuremath{\Lambda}CDM with
  high-redshift galaxy candidates}},
  \href{http://dx.doi.org/10.1038/s41550-023-01937-7}{\emph{Nature Astron.}
  {\bf 7} (2023) 731--735}, [\href{https://arxiv.org/abs/2208.01611}{{\tt
  2208.01611}}].

\bibitem{DES:2021cge}
{\scshape DES} collaboration, A.~Kov\'acs et~al., \emph{{The DES view of the
  Eridanus supervoid and the CMB cold spot}},
  \href{http://dx.doi.org/10.1093/mnras/stab3309}{\emph{Mon. Not. Roy. Astron.
  Soc.} {\bf 510} (2022) 216--229},
  [\href{https://arxiv.org/abs/2112.07699}{{\tt 2112.07699}}].

\bibitem{Bullock}
J.~S. {Bullock} and M.~{Boylan-Kolchin}, \emph{{Small-Scale Challenges to the
  {\ensuremath{\Lambda}}CDM Paradigm}},
  \href{http://dx.doi.org/10.1146/annurev-astro-091916-055313}{\emph{\araa}
  {\bf 55} (Aug., 2017) 343--387},
  [\href{https://arxiv.org/abs/1707.04256}{{\tt 1707.04256}}].

\bibitem{2012LRR....15...10F}
B.~{Famaey} and S.~S. {McGaugh}, \emph{{Modified Newtonian Dynamics (MOND):
  Observational Phenomenology and Relativistic Extensions}},
  \href{http://dx.doi.org/10.12942/lrr-2012-10}{\emph{Living Reviews in
  Relativity} {\bf 15} (Dec., 2012) 10},
  [\href{https://arxiv.org/abs/1112.3960}{{\tt 1112.3960}}].

\bibitem{Stahl:2023ccv}
C.~Stahl, Y.~Dubois, B.~Famaey, O.~Hahn, R.~Ibata, K.~Kraljic et~al.,
  \emph{{Hydrodynamical simulations of galaxy formation with non-Gaussian
  initial conditions}},
  \href{http://dx.doi.org/10.1088/1475-7516/2023/09/036}{\emph{JCAP} {\bf 09}
  (2023) 036}, [\href{https://arxiv.org/abs/2307.03300}{{\tt 2307.03300}}].

\bibitem{Stahl:2022did}
C.~Stahl, T.~Montandon, B.~Famaey, O.~Hahn and R.~Ibata, \emph{{Exploring the
  effects of primordial non-Gaussianity at galactic scales}},
  \href{http://dx.doi.org/10.1088/1475-7516/2023/01/024}{\emph{JCAP} {\bf 01}
  (2023) 024}, [\href{https://arxiv.org/abs/2209.15038}{{\tt 2209.15038}}].

\bibitem{Ezquiaga:2022qpw}
J.~M. Ezquiaga, J.~Garc\'\i{}a-Bellido and V.~Vennin, \emph{{Massive Galaxy
  Clusters Like El Gordo Hint at Primordial Quantum Diffusion}},
  \href{http://dx.doi.org/10.1103/PhysRevLett.130.121003}{\emph{Phys. Rev.
  Lett.} {\bf 130} (2023) 121003},
  [\href{https://arxiv.org/abs/2207.06317}{{\tt 2207.06317}}].

\bibitem{Biagetti:2022ode}
M.~Biagetti, G.~Franciolini and A.~Riotto, \emph{{High-redshift JWST
  Observations and Primordial Non-Gaussianity}},
  \href{http://dx.doi.org/10.3847/1538-4357/acb5ea}{\emph{Astrophys. J.} {\bf
  944} (2023) 113}, [\href{https://arxiv.org/abs/2210.04812}{{\tt
  2210.04812}}].

\bibitem{Mukhanov:2005sc}
V.~Mukhanov, \emph{{Physical Foundations of Cosmology}}.
\newblock Cambridge University Press, Oxford, 2005,
  \href{http://dx.doi.org/10.1017/CBO9780511790553}{10.1017/CBO9780511790553}.

\bibitem{Planck:2018vyg}
{\scshape Planck} collaboration, N.~Aghanim et~al., \emph{{Planck 2018 results.
  VI. Cosmological parameters}},
  \href{http://dx.doi.org/10.1051/0004-6361/201833910}{\emph{Astron.
  Astrophys.} {\bf 641} (2020) A6},
  [\href{https://arxiv.org/abs/1807.06209}{{\tt 1807.06209}}].

\bibitem{Tristram:2023haj}
M.~Tristram et~al., \emph{{Cosmological parameters derived from the final (PR4)
  Planck data release}},  \href{https://arxiv.org/abs/2309.10034}{{\tt
  2309.10034}}.

\bibitem{Chen:2018uul}
X.~Chen, G.~A. Palma, W.~Riquelme, B.~Scheihing~Hitschfeld and S.~Sypsas,
  \emph{{Landscape tomography through primordial non-Gaussianity}},
  \href{http://dx.doi.org/10.1103/PhysRevD.98.083528}{\emph{Phys. Rev. D} {\bf
  98} (2018) 083528}, [\href{https://arxiv.org/abs/1804.07315}{{\tt
  1804.07315}}].

\bibitem{Celoria:2021vjw}
M.~Celoria, P.~Creminelli, G.~Tambalo and V.~Yingcharoenrat, \emph{{Beyond
  perturbation theory in inflation}},
  \href{http://dx.doi.org/10.1088/1475-7516/2021/06/051}{\emph{JCAP} {\bf 06}
  (2021) 051}, [\href{https://arxiv.org/abs/2103.09244}{{\tt 2103.09244}}].

\bibitem{Gow:2022jfb}
A.~D. Gow, H.~Assadullahi, J.~H.~P. Jackson, K.~Koyama, V.~Vennin and D.~Wands,
  \emph{{Non-perturbative non-Gaussianity and primordial black holes}},
  \href{http://dx.doi.org/10.1209/0295-5075/acd417}{\emph{EPL} {\bf 142} (2023)
  49001}, [\href{https://arxiv.org/abs/2211.08348}{{\tt 2211.08348}}].

\bibitem{Cohen:2022clv}
T.~Cohen, D.~Green and A.~Premkumar, \emph{{Large deviations in the early
  Universe}}, \href{http://dx.doi.org/10.1103/PhysRevD.107.083501}{\emph{Phys.
  Rev. D} {\bf 107} (2023) 083501},
  [\href{https://arxiv.org/abs/2212.02535}{{\tt 2212.02535}}].

\bibitem{Hooshangi:2023kss}
S.~Hooshangi, M.~H. Namjoo and M.~Noorbala, \emph{{Tail diversity from
  inflation}},
  \href{http://dx.doi.org/10.1088/1475-7516/2023/09/023}{\emph{JCAP} {\bf 09}
  (2023) 023}, [\href{https://arxiv.org/abs/2305.19257}{{\tt 2305.19257}}].

\bibitem{Carr:2023tpt}
B.~Carr, S.~Clesse, J.~Garcia-Bellido, M.~Hawkins and F.~Kuhnel,
  \emph{{Observational Evidence for Primordial Black Holes: A Positivist
  Perspective}},  \href{https://arxiv.org/abs/2306.03903}{{\tt 2306.03903}}.

\bibitem{Clesse:2017bsw}
S.~Clesse and J.~Garc\'\i{}a-Bellido, \emph{{Seven Hints for Primordial Black
  Hole Dark Matter}},
  \href{http://dx.doi.org/10.1016/j.dark.2018.08.004}{\emph{Phys. Dark Univ.}
  {\bf 22} (2018) 137--146}, [\href{https://arxiv.org/abs/1711.10458}{{\tt
  1711.10458}}].

\bibitem{Green:2020jor}
A.~M. Green and B.~J. Kavanagh, \emph{{Primordial Black Holes as a dark matter
  candidate}}, \href{http://dx.doi.org/10.1088/1361-6471/abc534}{\emph{J. Phys.
  G} {\bf 48} (2021) 043001}, [\href{https://arxiv.org/abs/2007.10722}{{\tt
  2007.10722}}].

\bibitem{Caprini:2018mtu}
C.~Caprini and D.~G. Figueroa, \emph{{Cosmological Backgrounds of Gravitational
  Waves}}, \href{http://dx.doi.org/10.1088/1361-6382/aac608}{\emph{Class.
  Quant. Grav.} {\bf 35} (2018) 163001},
  [\href{https://arxiv.org/abs/1801.04268}{{\tt 1801.04268}}].

\bibitem{FrancoAbellan:2023sby}
G.~Franco~Abell\'an and G.~Facchinetti, \emph{{Minihalos as probes of the
  inflationary spectrum: accurate boost factor calculation and new CMB
  constraints}},
  \href{http://dx.doi.org/10.1088/1475-7516/2023/06/032}{\emph{JCAP} {\bf 06}
  (2023) 032}, [\href{https://arxiv.org/abs/2304.02996}{{\tt 2304.02996}}].

\bibitem{Bianchini:2022dqh}
F.~Bianchini and G.~Fabbian, \emph{{CMB spectral distortions revisited: A new
  take on $\mu$ distortions and primordial non-Gaussianities from FIRAS data}},
  \href{http://dx.doi.org/10.1103/PhysRevD.106.063527}{\emph{Phys. Rev. D} {\bf
  106} (2022) 063527}, [\href{https://arxiv.org/abs/2206.02762}{{\tt
  2206.02762}}].

\bibitem{LoVerde:2007ri}
M.~LoVerde, A.~Miller, S.~Shandera and L.~Verde, \emph{{Effects of
  Scale-Dependent Non-Gaussianity on Cosmological Structures}},
  \href{http://dx.doi.org/10.1088/1475-7516/2008/04/014}{\emph{JCAP} {\bf 04}
  (2008) 014}, [\href{https://arxiv.org/abs/0711.4126}{{\tt 0711.4126}}].

\bibitem{Pena:2022sdg}
G.~A. Pe\~na and G.~N. Candlish, \emph{{The large-scale structure from
  non-Gaussian primordial perturbations}},
  \href{http://dx.doi.org/10.1093/mnras/stac206}{\emph{Mon. Not. Roy. Astron.
  Soc.} {\bf 511} (2022) 2259--2273},
  [\href{https://arxiv.org/abs/2201.08842}{{\tt 2201.08842}}].

\bibitem{Anbajagane:2023wif}
D.~Anbajagane, C.~Chang, H.~Lee and M.~Gatti, \emph{{Primordial
  non-Gaussianities with weak lensing: Information on non-linear scales in the
  Ulagam full-sky simulations}},  \href{https://arxiv.org/abs/2310.02349}{{\tt
  2310.02349}}.

\bibitem{DAloisio:2011uez}
A.~D'Aloisio and P.~Natarajan, \emph{{The Effects of Primordial Non-Gaussianity
  on Giant-Arc Statistics: A Scale Dependent Example}},
  \href{http://dx.doi.org/10.22323/1.149.0025}{\emph{PoS} {\bf BASH11} (2011)
  025}, [\href{https://arxiv.org/abs/1202.0553}{{\tt 1202.0553}}].

\bibitem{Habouzit:2014hna}
M.~Habouzit, T.~Nishimichi, S.~Peirani, G.~A. Mamon, J.~Silk and J.~Chevallard,
  \emph{{Testing primordial non-Gaussianities on galactic scales at high
  redshift}}, \href{http://dx.doi.org/10.1093/mnrasl/slu145}{\emph{Mon. Not.
  Roy. Astron. Soc.} {\bf 445} (2014) 129},
  [\href{https://arxiv.org/abs/1407.8192}{{\tt 1407.8192}}].

\bibitem{Sabti:2020ser}
N.~Sabti, J.~B. Mu\~noz and D.~Blas, \emph{{First Constraints on Small-Scale
  Non-Gaussianity from UV Galaxy Luminosity Functions}},
  \href{http://dx.doi.org/10.1088/1475-7516/2021/01/010}{\emph{JCAP} {\bf 01}
  (2021) 010}, [\href{https://arxiv.org/abs/2009.01245}{{\tt 2009.01245}}].

\bibitem{Crociani:2008dt}
D.~Crociani, L.~Moscardini, M.~Viel and S.~Matarrese, \emph{{The effects of
  primordial non-Gaussianity on the cosmological reionization}},
  \href{http://dx.doi.org/10.1111/j.1365-2966.2008.14325.x}{\emph{Mon. Not.
  Roy. Astron. Soc.} {\bf 394} (2009) 133--141},
  [\href{https://arxiv.org/abs/0809.3909}{{\tt 0809.3909}}].

\bibitem{Chevallard:2014sxa}
J.~Chevallard, J.~Silk, T.~Nishimichi, M.~Habouzit, G.~A. Mamon and S.~Peirani,
  \emph{{Effect of primordial non-Gaussianities on the far-UV luminosity
  function of high-redshift galaxies: implications for cosmic reionization}},
  \href{http://dx.doi.org/10.1093/mnras/stu2280}{\emph{Mon. Not. Roy. Astron.
  Soc.} {\bf 446} (2015) 3235--3252},
  [\href{https://arxiv.org/abs/1410.7768}{{\tt 1410.7768}}].

\bibitem{Avila-Reese:2003cjm}
V.~Avila-Reese, P.~Colin, G.~Piccinelli and C.~Firmani, \emph{{The effects of
  non-Gaussian initial conditions on the structure and substructure of cold
  dark matter halos}}, \href{http://dx.doi.org/10.1086/378773}{\emph{Astrophys.
  J.} {\bf 598} (2003) 36--48},
  [\href{https://arxiv.org/abs/astro-ph/0306293}{{\tt astro-ph/0306293}}].

\bibitem{Cooray:2002dia}
A.~Cooray and R.~K. Sheth, \emph{{Halo Models of Large Scale Structure}},
  \href{http://dx.doi.org/10.1016/S0370-1573(02)00276-4}{\emph{Phys. Rept.}
  {\bf 372} (2002) 1--129}, [\href{https://arxiv.org/abs/astro-ph/0206508}{{\tt
  astro-ph/0206508}}].

\bibitem{Klasen:2015uma}
M.~Klasen, M.~Pohl and G.~Sigl, \emph{{Indirect and direct search for dark
  matter}}, \href{http://dx.doi.org/10.1016/j.ppnp.2015.07.001}{\emph{Prog.
  Part. Nucl. Phys.} {\bf 85} (2015) 1--32},
  [\href{https://arxiv.org/abs/1507.03800}{{\tt 1507.03800}}].

\bibitem{Angulo:2021kes}
R.~E. Angulo and O.~Hahn, \emph{{Large-scale dark matter simulations}},
  \href{https://arxiv.org/abs/2112.05165}{{\tt 2112.05165}}.

\bibitem{Navarro:1995iw}
J.~F. Navarro, C.~S. Frenk and S.~D.~M. White, \emph{{The Structure of cold
  dark matter halos}}, \href{http://dx.doi.org/10.1086/177173}{\emph{Astrophys.
  J.} {\bf 462} (1996) 563--575},
  [\href{https://arxiv.org/abs/astro-ph/9508025}{{\tt astro-ph/9508025}}].

\bibitem{Navarro:1996gj}
J.~F. Navarro, C.~S. Frenk and S.~D.~M. White, \emph{{A Universal density
  profile from hierarchical clustering}},
  \href{http://dx.doi.org/10.1086/304888}{\emph{Astrophys. J.} {\bf 490} (1997)
  493--508}, [\href{https://arxiv.org/abs/astro-ph/9611107}{{\tt
  astro-ph/9611107}}].

\bibitem{Jo}
J.~{Freundlich}, F.~{Jiang}, A.~{Dekel}, N.~{Cornuault}, O.~{Ginzburg},
  R.~{Koskas} et~al., \emph{{The Dekel-Zhao profile: a mass-dependent
  dark-matter density profile with flexible inner slope and analytic potential,
  velocity dispersion, and lensing properties}},
  \href{http://dx.doi.org/10.1093/mnras/staa2790}{\emph{\mnras} {\bf 499}
  (Dec., 2020) 2912--2933}, [\href{https://arxiv.org/abs/2004.08395}{{\tt
  2004.08395}}].

\bibitem{Korsaga_2023}
M.~Korsaga, B.~Famaey, J.~Freundlich, L.~Posti, R.~Ibata, C.~Boily et~al.,
  \emph{Disk galaxies are self-similar: The universality of the h i-to-halo
  mass ratio for isolated disks},
  \href{http://dx.doi.org/10.3847/2041-8213/ace364}{\emph{The Astrophysical
  Journal Letters} {\bf 952} (Aug., 2023) L41}.

\bibitem{Wang:2019ftp}
J.~Wang, S.~Bose, C.~S. Frenk, L.~Gao, A.~Jenkins, V.~Springel et~al.,
  \emph{{Universal structure of dark matter haloes over a mass range of 20
  orders of magnitude}},
  \href{http://dx.doi.org/10.1038/s41586-020-2642-9}{\emph{Nature} {\bf 585}
  (2020) 39--42}, [\href{https://arxiv.org/abs/1911.09720}{{\tt 1911.09720}}].

\bibitem{Ishiyama:2014uoa}
T.~Ishiyama, \emph{{Hierarchical Formation of Dark Matter Halos and the Free
  Streaming Scale}},
  \href{http://dx.doi.org/10.1088/0004-637X/788/1/27}{\emph{Astrophys. J.} {\bf
  788} (2014) 27}, [\href{https://arxiv.org/abs/1404.1650}{{\tt 1404.1650}}].

\bibitem{Colombi:2020xbv}
S.~Colombi, \emph{{Phase-space structure of protohalos: Vlasov versus
  Particle-Mesh}},
  \href{http://dx.doi.org/10.1051/0004-6361/202039719}{\emph{Astron.
  Astrophys.} {\bf 647} (2021) A66},
  [\href{https://arxiv.org/abs/2012.04409}{{\tt 2012.04409}}].

\bibitem{Delos_2022}
M.~S. Delos and S.~D.~M. White, \emph{Inner cusps of the first dark matter
  haloes: formation and survival in a cosmological context},
  \href{http://dx.doi.org/10.1093/mnras/stac3373}{\emph{Monthly Notices of the
  Royal Astronomical Society} {\bf 518} (Nov., 2022) 3509–3532}.

\bibitem{Wang:2008un}
J.~Wang and S.~D.~M. White, \emph{{Are mergers responsible for universal halo
  properties?}},
  \href{http://dx.doi.org/10.1111/j.1365-2966.2009.14755.x}{\emph{Mon. Not.
  Roy. Astron. Soc.} {\bf 396} (2009) 709},
  [\href{https://arxiv.org/abs/0809.1322}{{\tt 0809.1322}}].

\bibitem{Ludlow:2013bd}
A.~D. Ludlow, J.~F. Navarro, M.~Boylan-Kolchin, P.~E. Bett, R.~E. Angulo, M.~Li
  et~al., \emph{{The Mass Profile and Accretion History of Cold Dark Matter
  Halos}}, \href{http://dx.doi.org/10.1093/mnras/stt526}{\emph{Mon. Not. Roy.
  Astron. Soc.} {\bf 432} (2013) 1103},
  [\href{https://arxiv.org/abs/1302.0288}{{\tt 1302.0288}}].

\bibitem{2010arXiv1010.2539D}
N.~{Dalal}, Y.~{Lithwick} and M.~{Kuhlen}, \emph{{The Origin of Dark Matter
  Halo Profiles}},
  \href{http://dx.doi.org/10.48550/arXiv.1010.2539}{\emph{arXiv e-prints}
  (Oct., 2010) arXiv:1010.2539}, [\href{https://arxiv.org/abs/1010.2539}{{\tt
  1010.2539}}].

\bibitem{2013MNRAS.430..121P}
A.~{Pontzen} and F.~{Governato}, \emph{{Conserved actions, maximum entropy and
  dark matter haloes}},
  \href{http://dx.doi.org/10.1093/mnras/sts529}{\emph{\mnras} {\bf 430} (Mar.,
  2013) 121--133}, [\href{https://arxiv.org/abs/1210.1849}{{\tt 1210.1849}}].

\bibitem{Wagner:2020opz}
J.~Wagner, \emph{{Cosmic structures from a mathematical perspective 1. Dark
  matter halo mass density profiles}},
  \href{http://dx.doi.org/10.1007/s10714-020-02715-w}{\emph{Gen. Rel. Grav.}
  {\bf 52} (2020) 61}, [\href{https://arxiv.org/abs/2002.00960}{{\tt
  2002.00960}}].

\bibitem{Baushev:2013ida}
A.~N. Baushev, \emph{{The real and apparent convergence of N-body simulations
  of the dark matter structures: Is the
  Navarro\textendash{}Frenk\textendash{}White profile real?}},
  \href{http://dx.doi.org/10.1016/j.astropartphys.2014.07.012}{\emph{Astropart.
  Phys.} {\bf 62} (2015) 47--53}, [\href{https://arxiv.org/abs/1312.0314}{{\tt
  1312.0314}}].

\bibitem{Brown:2020lxj}
S.~T. Brown, I.~G. McCarthy, B.~Diemer, A.~S. Font, S.~G. Stafford and
  S.~Pfeifer, \emph{{Connecting the structure of dark matter haloes to the
  primordial power spectrum}},  \href{https://arxiv.org/abs/2005.12933}{{\tt
  2005.12933}}.

\bibitem{Fedeli:2009mt}
C.~Fedeli and L.~Moscardini, \emph{{Cosmic shear statistics in cosmologies with
  non-Gaussian initial conditions}},
  \href{http://dx.doi.org/10.1111/j.1365-2966.2010.16500.x}{\emph{Mon. Not.
  Roy. Astron. Soc.} {\bf 405} (2010) 681--694},
  [\href{https://arxiv.org/abs/0912.4112}{{\tt 0912.4112}}].

\bibitem{Smith:2010fh}
R.~E. Smith, V.~Desjacques and L.~Marian, \emph{{Nonlinear clustering in models
  with primordial non-Gaussianity: the halo model approach}},
  \href{http://dx.doi.org/10.1103/PhysRevD.83.043526}{\emph{Phys. Rev. D} {\bf
  83} (2011) 043526}, [\href{https://arxiv.org/abs/1009.5085}{{\tt
  1009.5085}}].

\bibitem{Figueroa:2012ws}
D.~G. Figueroa, E.~Sefusatti, A.~Riotto and F.~Vernizzi, \emph{{The Effect of
  Local non-Gaussianity on the Matter Bispectrum at Small Scales}},
  \href{http://dx.doi.org/10.1088/1475-7516/2012/08/036}{\emph{JCAP} {\bf 08}
  (2012) 036}, [\href{https://arxiv.org/abs/1205.2015}{{\tt 1205.2015}}].

\bibitem{MoradinezhadDizgah:2013rkr}
A.~Moradinezhad~Dizgah, S.~Dodelson and A.~Riotto, \emph{{Imprint of Primordial
  Non-Gaussianity on Dark Matter Halo Profiles}},
  \href{http://dx.doi.org/10.1103/PhysRevD.88.063513}{\emph{Phys. Rev. D} {\bf
  88} (2013) 063513}, [\href{https://arxiv.org/abs/1307.2632}{{\tt
  1307.2632}}].

\bibitem{Springel:2020plp}
V.~Springel, R.~Pakmor, O.~Zier and M.~Reinecke, \emph{{Simulating cosmic
  structure formation with the gadget-4 code}},
  \href{http://dx.doi.org/10.1093/mnras/stab1855}{\emph{Mon. Not. Roy. Astron.
  Soc.} {\bf 506} (2021) 2871--2949},
  [\href{https://arxiv.org/abs/2010.03567}{{\tt 2010.03567}}].

\bibitem{Teyssier:2001cp}
R.~Teyssier, \emph{{Cosmological hydrodynamics with adaptive mesh refinement: a
  new high resolution code called ramses}},
  \href{http://dx.doi.org/10.1051/0004-6361:20011817}{\emph{Astron. Astrophys.}
  {\bf 385} (2002) 337--364},
  [\href{https://arxiv.org/abs/astro-ph/0111367}{{\tt astro-ph/0111367}}].

\bibitem{Dubois:2014lxa}
Y.~Dubois et~al., \emph{{Dancing in the dark: galactic properties trace spin
  swings along the cosmic web}},
  \href{http://dx.doi.org/10.1093/mnras/stu1227}{\emph{Mon. Not. Roy. Astron.
  Soc.} {\bf 444} (2014) 1453--1468},
  [\href{https://arxiv.org/abs/1402.1165}{{\tt 1402.1165}}].

\bibitem{Dubois_2016}
Y.~Dubois, S.~Peirani, C.~Pichon, J.~Devriendt, R.~Gavazzi, C.~Welker et~al.,
  \emph{The horizon-{AGN} simulation: morphological diversity of galaxies
  promoted by {AGN} feedback},
  \href{http://dx.doi.org/10.1093/mnras/stw2265}{\emph{Monthly Notices of the
  Royal Astronomical Society} {\bf 463} (sep, 2016) 3948--3964}.

\bibitem{Planck:2019kim}
{\scshape Planck} collaboration, Y.~Akrami et~al., \emph{{Planck 2018 results.
  IX. Constraints on primordial non-Gaussianity}},
  \href{http://dx.doi.org/10.1051/0004-6361/201935891}{\emph{Astron.
  Astrophys.} {\bf 641} (2020) A9},
  [\href{https://arxiv.org/abs/1905.05697}{{\tt 1905.05697}}].

\bibitem{Khoury:2008wj}
J.~Khoury and F.~Piazza, \emph{{Rapidly-Varying Speed of Sound, Scale
  Invariance and Non-Gaussian Signatures}},
  \href{http://dx.doi.org/10.1088/1475-7516/2009/07/026}{\emph{JCAP} {\bf 07}
  (2009) 026}, [\href{https://arxiv.org/abs/0811.3633}{{\tt 0811.3633}}].

\bibitem{Riotto:2010nh}
A.~Riotto and M.~S. Sloth, \emph{{Strongly Scale-dependent Non-Gaussianity}},
  \href{http://dx.doi.org/10.1103/PhysRevD.83.041301}{\emph{Phys. Rev. D} {\bf
  83} (2011) 041301}, [\href{https://arxiv.org/abs/1009.3020}{{\tt
  1009.3020}}].

\bibitem{Byrnes:2011gh}
C.~T. Byrnes, K.~Enqvist, S.~Nurmi and T.~Takahashi, \emph{{Strongly
  scale-dependent polyspectra from curvaton self-interactions}},
  \href{http://dx.doi.org/10.1088/1475-7516/2011/11/011}{\emph{JCAP} {\bf 11}
  (2011) 011}, [\href{https://arxiv.org/abs/1108.2708}{{\tt 1108.2708}}].

\bibitem{Jackson:2023obv}
J.~H.~P. Jackson, H.~Assadullahi, A.~D. Gow, K.~Koyama, V.~Vennin and D.~Wands,
  \emph{{The separate-universe approach and sudden transitions during
  inflation}},  \href{https://arxiv.org/abs/2311.03281}{{\tt 2311.03281}}.

\bibitem{Pinol:2023oux}
L.~Pinol, S.~Renaux-Petel and D.~Werth, \emph{{The Cosmological Flow: A
  Systematic Approach to Primordial Correlators}},
  \href{https://arxiv.org/abs/2312.06559}{{\tt 2312.06559}}.

\bibitem{Michaux:2020yis}
M.~Michaux, O.~Hahn, C.~Rampf and R.~E. Angulo, \emph{{Accurate initial
  conditions for cosmological N-body simulations: Minimizing truncation and
  discreteness errors}},
  \href{http://dx.doi.org/10.1093/mnras/staa3149}{\emph{Mon. Not. Roy. Astron.
  Soc.} {\bf 500} (2020) 663--683},
  [\href{https://arxiv.org/abs/2008.09588}{{\tt 2008.09588}}].

\bibitem{Springel:2000qu}
V.~Springel, S.~D.~M. White, G.~Tormen and G.~Kauffmann, \emph{{Populating a
  cluster of galaxies. 1. Results at z = 0}},
  \href{http://dx.doi.org/10.1046/j.1365-8711.2001.04912.x}{\emph{Mon. Not.
  Roy. Astron. Soc.} {\bf 328} (2001) 726},
  [\href{https://arxiv.org/abs/astro-ph/0012055}{{\tt astro-ph/0012055}}].

\bibitem{Aubert:2004mu}
D.~Aubert, C.~Pichon and S.~Colombi, \emph{{The Origin and implications of dark
  matter anisotropic cosmic infall on \textasciitilde{} L(*) halos}},
  \href{http://dx.doi.org/10.1111/j.1365-2966.2004.07883.x}{\emph{Mon. Not.
  Roy. Astron. Soc.} {\bf 352} (2004) 376},
  [\href{https://arxiv.org/abs/astro-ph/0402405}{{\tt astro-ph/0402405}}].

\bibitem{2013ApJ...762..109B}
P.~S. {Behroozi}, R.~H. {Wechsler} and H.-Y. {Wu}, \emph{{The ROCKSTAR
  Phase-space Temporal Halo Finder and the Velocity Offsets of Cluster Cores}},
  \href{http://dx.doi.org/10.1088/0004-637X/762/2/109}{\emph{\apj} {\bf 762}
  (Jan., 2013) 109}, [\href{https://arxiv.org/abs/1110.4372}{{\tt 1110.4372}}].

\bibitem{Ludlow:2016ifl}
A.~D. Ludlow, S.~Bose, R.~E. Angulo, L.~Wang, W.~A. Hellwing, J.~F. Navarro
  et~al., \emph{{The mass\textendash{}concentration\textendash{}redshift
  relation of cold and warm dark matter haloes}},
  \href{http://dx.doi.org/10.1093/mnras/stw1046}{\emph{Mon. Not. Roy. Astron.
  Soc.} {\bf 460} (2016) 1214--1232},
  [\href{https://arxiv.org/abs/1601.02624}{{\tt 1601.02624}}].

\bibitem{Diemer:2017bwl}
B.~Diemer, \emph{{COLOSSUS: A python toolkit for cosmology, large-scale
  structure, and dark matter halos}},
  \href{http://dx.doi.org/10.3847/1538-4365/aaee8c}{\emph{Astrophys. J. Suppl.}
  {\bf 239} (2018) 35}, [\href{https://arxiv.org/abs/1712.04512}{{\tt
  1712.04512}}].

\bibitem{Balmes:2013hqa}
I.~Balm\`es, Y.~Rasera, P.-S. Corasaniti and J.-M. Alimi, \emph{{Imprints of
  dark energy on cosmic structure formation \textendash{} III. Sparsity of dark
  matter halo profiles}},
  \href{http://dx.doi.org/10.1093/mnras/stt2050}{\emph{Mon. Not. Roy. Astron.
  Soc.} {\bf 437} (2014) 2328--2339},
  [\href{https://arxiv.org/abs/1307.2922}{{\tt 1307.2922}}].

\bibitem{Corasaniti:2017yts}
P.~S. Corasaniti, S.~Ettori, Y.~Rasera, M.~Sereno, S.~Amodeo, M.~A. Breton
  et~al., \emph{{Probing Cosmology with Dark Matter Halo Sparsity Using X-ray
  Cluster Mass Measurements}},
  \href{http://dx.doi.org/10.3847/1538-4357/aaccdf}{\emph{Astrophys. J.} {\bf
  862} (2018) 40}, [\href{https://arxiv.org/abs/1711.00480}{{\tt 1711.00480}}].

\bibitem{Corasaniti_2019}
P.~S. Corasaniti and Y.~Rasera, \emph{Average dark matter halo sparsity
  relations as consistency check of mass estimates in galaxy cluster samples},
  \href{http://dx.doi.org/10.1093/mnras/stz1579}{\emph{Monthly Notices of the
  Royal Astronomical Society} {\bf 487} (June, 2019) 4382–4392}.

\bibitem{Corasaniti_2020}
P.~S. Corasaniti, C.~Giocoli and M.~Baldi, \emph{Dark matter halo sparsity of
  modified gravity scenarios},
  \href{http://dx.doi.org/10.1103/physrevd.102.043501}{\emph{Physical Review D}
  {\bf 102} (Aug., 2020) }.

\bibitem{Corasaniti_2022}
P.~S. Corasaniti, A.~M.~C. Le Brun, T.~R.~G. Richardson, Y.~Rasera, S.~Ettori,
  M.~Arnaud et~al., \emph{Forecasting cosmological parameter constraints using
  multiple sparsity measurements as tracers of the mass profiles of dark matter
  haloes}, \href{http://dx.doi.org/10.1093/mnras/stac2196}{\emph{Monthly
  Notices of the Royal Astronomical Society} {\bf 516} (Aug., 2022) 437–452}.

\bibitem{Richardson:2021yvu}
T.~R.~G. Richardson and P.~S. Corasaniti, \emph{{Timing the last major merger
  of galaxy clusters with large halo sparsity}},
  \href{http://dx.doi.org/10.1093/mnras/stac1241}{\emph{Mon. Not. Roy. Astron.
  Soc.} {\bf 513} (2022) 4951--4967},
  [\href{https://arxiv.org/abs/2112.04926}{{\tt 2112.04926}}].

\bibitem{Richardson:2022cvd}
T.~R.~G. Richardson and P.-S. Corasaniti, \emph{{A non-parametric approach to
  the relation between the halo mass function and internal dark matter
  structure of haloes}},
  \href{http://dx.doi.org/10.1051/0004-6361/202245622}{\emph{Astron.
  Astrophys.} {\bf 674} (2023) A173},
  [\href{https://arxiv.org/abs/2212.03233}{{\tt 2212.03233}}].

\bibitem{Tamara}
T.~R.~G. Richardson, \emph{{Implications cosmologiques et astrophysiques de la
  structure interne des halos de matière noire}}.
\newblock PhD thesis, Observatoire de Paris, 2023.

\bibitem{Einasto}
J.~{Einasto}, \emph{{On the Construction of a Composite Model for the Galaxy
  and on the Determination of the System of Galactic Parameters}}, {\emph{Trudy
  Astrofizicheskogo Instituta Alma-Ata} {\bf 5} (Jan., 1965) 87--100}.

\bibitem{Jaffe}
W.~{Jaffe}, \emph{{A simple model for the distribution of light in spherical
  galaxies.}}, \href{http://dx.doi.org/10.1093/mnras/202.4.995}{\emph{\mnras}
  {\bf 202} (Mar., 1983) 995--999}.

\bibitem{Hernquist}
L.~{Hernquist}, \emph{{An Analytical Model for Spherical Galaxies and Bulges}},
  \href{http://dx.doi.org/10.1086/168845}{\emph{\apj} {\bf 356} (June, 1990)
  359}.

\bibitem{Dehnen}
W.~{Dehnen}, \emph{{A Family of Potential-Density Pairs for Spherical Galaxies
  and Bulges}}, \href{http://dx.doi.org/10.1093/mnras/265.1.250}{\emph{\mnras}
  {\bf 265} (Nov., 1993) 250}.

\bibitem{Evans}
N.~W. {Evans}, \emph{{The power-law galaxies.}},
  \href{http://dx.doi.org/10.1093/mnras/267.2.333}{\emph{\mnras} {\bf 267}
  (Mar., 1994) 333--360}.

\bibitem{Tremaine}
S.~{Tremaine}, D.~O. {Richstone}, Y.-I. {Byun}, A.~{Dressler}, S.~M. {Faber},
  C.~{Grillmair} et~al., \emph{{A Family of Models for Spherical Stellar
  Systems}}, \href{http://dx.doi.org/10.1086/116883}{\emph{\aj} {\bf 107}
  (Feb., 1994) 634}, [\href{https://arxiv.org/abs/astro-ph/9309044}{{\tt
  astro-ph/9309044}}].

\bibitem{Burkert}
A.~{Burkert}, \emph{{The Structure of Dark Matter Halos in Dwarf Galaxies}},
  \href{http://dx.doi.org/10.1086/309560}{\emph{\apjl} {\bf 447} (July, 1995)
  L25--L28}, [\href{https://arxiv.org/abs/astro-ph/9504041}{{\tt
  astro-ph/9504041}}].

\bibitem{Merritt}
D.~{Merritt}, A.~W. {Graham}, B.~{Moore}, J.~{Diemand} and B.~{Terzi{\'c}},
  \emph{{Empirical Models for Dark Matter Halos. I. Nonparametric Construction
  of Density Profiles and Comparison with Parametric Models}},
  \href{http://dx.doi.org/10.1086/508988}{\emph{\aj} {\bf 132} (Dec., 2006)
  2685--2700}, [\href{https://arxiv.org/abs/astro-ph/0509417}{{\tt
  astro-ph/0509417}}].

\bibitem{2010arXiv1005.0411C}
D.~{Coe}, \emph{{Dark Matter Halo Mass Profiles}},
  \href{http://dx.doi.org/10.48550/arXiv.1005.0411}{\emph{arXiv e-prints} (May,
  2010) arXiv:1005.0411}, [\href{https://arxiv.org/abs/1005.0411}{{\tt
  1005.0411}}].

\bibitem{Read}
J.~I. {Read}, O.~{Agertz} and M.~L.~M. {Collins}, \emph{{Dark matter cores all
  the way down}}, \href{http://dx.doi.org/10.1093/mnras/stw713}{\emph{\mnras}
  {\bf 459} (July, 2016) 2573--2590},
  [\href{https://arxiv.org/abs/1508.04143}{{\tt 1508.04143}}].

\bibitem{Lazar}
A.~{Lazar}, J.~S. {Bullock}, M.~{Boylan-Kolchin}, T.~K. {Chan}, P.~F.
  {Hopkins}, A.~S. {Graus} et~al., \emph{{A dark matter profile to model
  diverse feedback-induced core sizes of {\ensuremath{\Lambda}}CDM haloes}},
  \href{http://dx.doi.org/10.1093/mnras/staa2101}{\emph{\mnras} {\bf 497}
  (Sept., 2020) 2393--2417}, [\href{https://arxiv.org/abs/2004.10817}{{\tt
  2004.10817}}].

\bibitem{Zhao}
H.~{Zhao}, \emph{{Analytical models for galactic nuclei}},
  \href{http://dx.doi.org/10.1093/mnras/278.2.488}{\emph{\mnras} {\bf 278}
  (Jan., 1996) 488--496}, [\href{https://arxiv.org/abs/astro-ph/9509122}{{\tt
  astro-ph/9509122}}].

\bibitem{Begue:2017lcw}
D.~B\'egu\'e, C.~Stahl and S.-S. Xue, \emph{{A model of interacting dark fluids
  tested with supernovae and Baryon Acoustic Oscillations data}},
  \href{http://dx.doi.org/10.1016/j.nuclphysb.2019.01.001}{\emph{Nucl. Phys. B}
  {\bf 940} (2019) 312--320}, [\href{https://arxiv.org/abs/1702.03185}{{\tt
  1702.03185}}].

\bibitem{Ghari}
A.~{Ghari}, B.~{Famaey}, C.~{Laporte} and H.~{Haghi}, \emph{{Dark matter-baryon
  scaling relations from Einasto halo fits to SPARC galaxy rotation curves}},
  \href{http://dx.doi.org/10.1051/0004-6361/201834661}{\emph{\aap} {\bf 623}
  (Mar., 2019) A123}, [\href{https://arxiv.org/abs/1811.06554}{{\tt
  1811.06554}}].

\bibitem{Gao:2007gh}
L.~Gao, J.~F. Navarro, S.~Cole, C.~Frenk, S.~D.~M. White, V.~Springel et~al.,
  \emph{{The redshift dependence of the structure of massive LCDM halos}},
  \href{http://dx.doi.org/10.1111/j.1365-2966.2008.13277.x}{\emph{Mon. Not.
  Roy. Astron. Soc.} {\bf 387} (2008) 536},
  [\href{https://arxiv.org/abs/0711.0746}{{\tt 0711.0746}}].

\bibitem{Dutton:2014xda}
A.~A. Dutton and A.~V. Macci\`o, \emph{{Cold dark matter haloes in the Planck
  era: evolution of structural parameters for Einasto and NFW profiles}},
  \href{http://dx.doi.org/10.1093/mnras/stu742}{\emph{Mon. Not. Roy. Astron.
  Soc.} {\bf 441} (2014) 3359--3374},
  [\href{https://arxiv.org/abs/1402.7073}{{\tt 1402.7073}}].

\bibitem{Ludlow:2016qow}
A.~D. Ludlow and R.~E. Angulo, \emph{{Einasto Profiles and the Dark Matter
  Power Spectrum}}, \href{http://dx.doi.org/10.1093/mnrasl/slw216}{\emph{Mon.
  Not. Roy. Astron. Soc.} {\bf 465} (2017) L84--L88},
  [\href{https://arxiv.org/abs/1610.04620}{{\tt 1610.04620}}].

\bibitem{Lucie-Smith:2022uvv}
L.~Lucie-Smith, H.~V. Peiris, A.~Pontzen, B.~Nord, J.~Thiyagalingam and
  D.~Piras, \emph{{Discovering the building blocks of dark matter halo density
  profiles with neural networks}},
  \href{http://dx.doi.org/10.1103/PhysRevD.105.103533}{\emph{Phys. Rev. D} {\bf
  105} (2022) 103533}, [\href{https://arxiv.org/abs/2203.08827}{{\tt
  2203.08827}}].

\bibitem{Lucie-Smith:2023kue}
L.~Lucie-Smith, H.~V. Peiris and A.~Pontzen, \emph{{Explaining dark matter halo
  density profiles with neural networks}},
  \href{https://arxiv.org/abs/2305.03077}{{\tt 2305.03077}}.

\bibitem{WT1}
W.~{Tenachi}, R.~{Ibata} and F.~I. {Diakogiannis}, \emph{{Deep Symbolic
  Regression for Physics Guided by Units Constraints: Toward the Automated
  Discovery of Physical Laws}},
  \href{http://dx.doi.org/10.3847/1538-4357/ad014c}{\emph{\apj} {\bf 959}
  (Dec., 2023) 99}, [\href{https://arxiv.org/abs/2303.03192}{{\tt
  2303.03192}}].

\bibitem{WT2}
W.~{Tenachi}, R.~{Ibata}, T.~L. {Fran{\c{c}}ois} and F.~I. {Diakogiannis},
  \emph{{Class Symbolic Regression: Gotta Fit 'Em All}},
  \href{http://dx.doi.org/10.48550/arXiv.2312.01816}{\emph{arXiv e-prints}
  (Dec., 2023) arXiv:2312.01816}, [\href{https://arxiv.org/abs/2312.01816}{{\tt
  2312.01816}}].

\bibitem{Turk:2010ah}
M.~J. {Turk}, B.~D. {Smith}, J.~S. {Oishi}, S.~{Skory}, S.~W. {Skillman},
  T.~{Abel} et~al., \emph{{yt: A Multi-code Analysis Toolkit for Astrophysical
  Simulation Data}},
  \href{http://dx.doi.org/10.1088/0067-0049/192/1/9}{\emph{\apjs} {\bf 192}
  (Jan., 2011) 9}, [\href{https://arxiv.org/abs/1011.3514}{{\tt 1011.3514}}].

\bibitem{Perez:2007emg}
F.~Perez and B.~E. Granger, \emph{{IPython: A System for Interactive Scientific
  Computing}}, \href{http://dx.doi.org/10.1109/MCSE.2007.53}{\emph{Comput. Sci.
  Eng.} {\bf 9} (2007) 21--29}.

\bibitem{Hunter:2007ouj}
J.~D. Hunter, \emph{{Matplotlib: A 2D Graphics Environment}},
  \href{http://dx.doi.org/10.1109/MCSE.2007.55}{\emph{Comput. Sci. Eng.} {\bf
  9} (2007) 90--95}.

\bibitem{vanderWalt:2011bqk}
S.~van~der Walt, S.~C. Colbert and G.~Varoquaux, \emph{{The NumPy Array: A
  Structure for Efficient Numerical Computation}},
  \href{http://dx.doi.org/10.1109/MCSE.2011.37}{\emph{Comput. Sci. Eng.} {\bf
  13} (2011) 22--30}, [\href{https://arxiv.org/abs/1102.1523}{{\tt
  1102.1523}}].

\bibitem{Virtanen:2019joe}
P.~Virtanen et~al., \emph{{SciPy 1.0--Fundamental Algorithms for Scientific
  Computing in Python}},
  \href{http://dx.doi.org/10.1038/s41592-019-0686-2}{\emph{Nature Meth.} {\bf
  17} (2020) 261}, [\href{https://arxiv.org/abs/1907.10121}{{\tt 1907.10121}}].

\bibitem{berthoud}
F.~Berthoud, B.~Bzeznik, N.~Gibelin, M.~Laurens, C.~Bonamy, M.~Morel et~al.,
  \emph{{Estimation de l'empreinte carbone d'une heure.coeur de calcul}},
  research report, {UGA - Universit{\'e} Grenoble Alpes ; CNRS ; INP Grenoble ;
  INRIA}, Apr., 2020.

\end{thebibliography}\endgroup

\end{document}